\documentclass[aps,pre,twocolumn,superscriptaddress]{revtex4-2}

\usepackage{graphicx}
\usepackage{amsmath}
\usepackage{microtype}
\usepackage{array}
\usepackage{multirow}
\usepackage{dcolumn}
\usepackage{hyperref}
\hypersetup{breaklinks=true}
\usepackage{float}

\begin{document}

\title{Structure-based approach to identifying small sets of driver nodes in biological networks}

\date{\today}

\author{Eli Newby}
\email[]{eyn5@psu.edu}
\affiliation{Department of Physics, The Pennsylvania State University, University Park, PA 16802}
\author{Jorge G\'{o}mez Tejeda Za\~{n}udo}
\email[]{jgtz@broadinstitute.org}
\affiliation{Eli and Edythe L. Broad Institute of MIT and Harvard, Cambridge, MA 02142}
\affiliation{Department of Medical Oncology, Dana-Farber Cancer Institute, Harvard Medical School, Boston, MA 02115}
\author{R\'{e}ka Albert}
\email[]{rza1@psu.edu}
\affiliation{Department of Physics, The Pennsylvania State University, University Park, PA 16802}
\affiliation{Department of Biology, The Pennsylvania State University, University Park, PA 16802}

\begin{abstract}
In network control theory, driving all the nodes in the Feedback Vertex Set (FVS) forces the network into one of its attractors (long-term dynamic behaviors). The FVS is often composed of more nodes than can be realistically manipulated in a system; for example, only up to three nodes can be controlled in intracellular networks, while their FVS may contain more than 10 nodes. Thus, we developed an approach to rank subsets of the FVS on Boolean models of intracellular networks using topological, dynamics-independent measures. We investigated the use of seven topological prediction measures sorted into three categories -- centrality measures, propagation measures, and cycle-based measures. Using each measure every subset was ranked and then evaluated against two dynamics-based metrics that measure the ability of interventions to drive the system towards or away from its attractors: \textit{To Control} and \textit{Away Control}. After examining an array of biological networks, we found that the FVS subsets that ranked in the top according to the propagation metrics can most effectively control the network. This result was independently corroborated on a second array of different Boolean models of biological networks. Consequently, overriding the entire FVS is not required to drive a biological network to one of its attractors, and this method provides a way to reliably identify effective FVS subsets without knowledge of the network's dynamics.
\end{abstract}
\maketitle

\section{\label{sec:Intro}Introduction}
Complex systems are made up of many interacting components; they cannot be understood by looking at each component individually and instead should be studied as collectives \cite{Boccara,Schweitzer,Sayama,Shalizi}. Power grids, telecommunication systems, ecological and social systems, the nervous system, and cells are all examples of complex systems. The goal of studying these complex systems is to understand how their dynamic behavior emerges from the interactions of their entities.

Networks are one of the models used to encode the intricate interactions that underlie complex systems \cite{Newman,Watts}. A network representation uses nodes to denote the components of the system and edges to represent their interactions. In some contexts, the only information given about a complex system is the network representation with its nodes and edges, which can be studied to gain valuable information \cite{Milo, Ravasz, Alm}. In other cases, dynamics are also specified on the network to describe the processes that take place in the system, for example the propagation of current (in power grids), information (in telecommunication or the nervous system), disease (in ecological or social systems) \cite{Vespignani}. These dynamic processes are described by assigning a time-dependent variable (continuous or discrete) to each node. The time evolution of this node variable is based on how its incoming edges affect it, which produces a node update equation. The set of node update equations are important to grasp how the system's state changes in time.

In addition to studying complex systems' autonomous dynamics, understanding how to influence these dynamics to reach important states is of great interest. Network control has many theoretical and practical facets to determine the best course of action for controlling a complex system \cite{Liu,Liu2,Ruths,Gates,Cornelius,Wang,Nacher,Mochizuki,Zanudo}. Different control methods are applicable based on the information available (e.g., whether dynamic information is available or not) and on the goal (e.g., which state we want to reach). From here, control methods aim to find a set of elements and the corresponding actions on them required to attain the desired objective and then evaluate the practicality of this set.

We can apply these ideas to biology by representing biological systems as networks and modeling biological processes as dynamic information propagation on these networks. In biological networks at the cellular level, nodes represent biological macromolecules and edges represent the interactions between these macromolecules. These interactions underlie cell phenotypes and behaviors such as movement, cell division, and programmed cell death. Dynamical models on a cellular network are used to describe the signal transduction, gene regulation and metabolic processes of the biological system \cite{Abou,Wynn,Zanudo4,Alon,Tyson}. Thus, the models' dynamics must capture the long-term steady states -- attractors -- of the cell (e.g., cell types).

When attempting to control complex biological systems, reaching an important attractor is often the main objective. Attractors are sections of the state space that the system can enter but cannot exit. They can be either simple attractors (e.g., fixed points) or complex attractors (e.g., stable oscillations). Along with each attractor, it is valuable to identify the attractor's basin of attraction -- the region of state space from which every initial state will lead to that specific attractor.

\begin{figure}
\includegraphics[width=\linewidth]{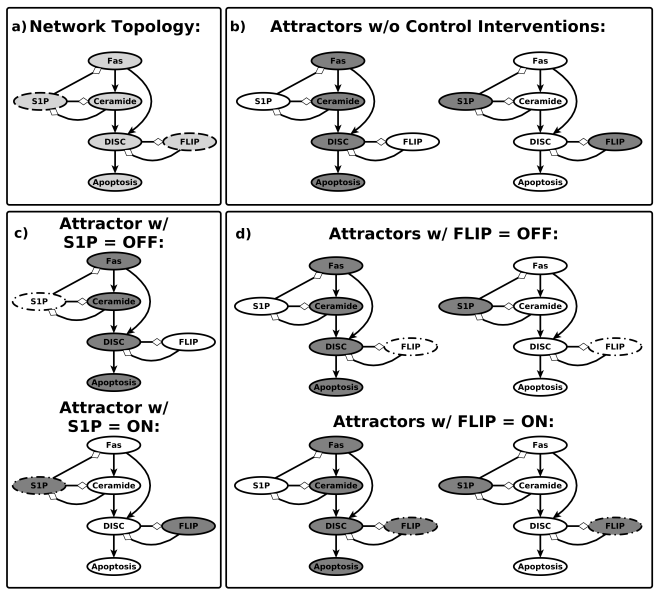}
\caption{\label{fig1} (a) A simplified model of the T-LGL network. A minimal FVS of this network is \{S1P, FLIP\}, which is highlighted with dashed outlines. (b) The two attractors of this network (white represents OFF and dark grey represents ON). The attractors represent the two cell states of the network, one where the cell commits to cell death (apoptosis is ON) and another where the cell survives (apoptosis is OFF). We investigate how fixing only one of the nodes of the FVS (dash-dotted outline) affects the network's dynamics. (c) Fixing S1P to the OFF state drives the system to the apoptosis attractor, and fixing S1P to the ON state drives the system to the survival attractor. (d) Conversely, fixing FLIP to its OFF state or fixing FLIP to its ON state preserves both attractor, thus both interventions fail to drive the system. The topological difference between these two intervention targets is of interest to generalize why any one subset would outperform another subset.}
\end{figure}

As an example, the cell network in Fig. \ref{fig1} represents a simplified version of the signaling network that underlies T cell large granular lymphocyte (T-LGL) leukemia \cite{Saadatpour}. There are two attractors, which represent two cell fates: a state of commitment to cell death (apoptosis) and a state of cell survival, which is pathological in this case. Because this is a cancer network, it would be beneficial to understand how a cell becomes cancerous (i.e., which states are in the basin of attraction of the cancerous survival state) and how to induce cell death (i.e., how to drive the system to the apoptosis attractor). This analysis helps to understand which specific molecules are biologically important for T-LGL leukemia and leads to predictions of therapeutic interventions.

One method for driving a network to any of its natural attractors (attractor control) depends on the connection between the feedback vertex set (FVS) and the attractors of the system \cite{Mochizuki, Zanudo}. The FVS is a set of nodes in the network that contains nodes in every cycle of the network. In \cite{Mochizuki}, Mochizuki et al. proved that, given a system governed by a general class of nonlinear dynamics on an underlying network structure, driving the state of every node in the FVS to its corresponding state in a target attractor is guaranteed to drive the system to this target attractor \cite{Zanudo}.

For our purposes, we consider the minimal FVS, which is the smallest set of nodes that accomplishes this task. This set is not necessarily unique. In Fig. \ref{fig1}(a), we label a possible minimal FVS in with dashed outlines; however, Ceramide is part of the same cycles as S1P, so it could be part of the minimal FVS instead of S1P. Similarly, DISC could replace FLIP in the minimal FVS. Figure \ref{fig1} focuses on the minimal FVS of S1P and FLIP, but all minimal FVSs are probed in our broader analysis.

To better understand the connection between the FVS and the attractors of the system, it is useful to relate it to previous work connecting cycles in the network structure to the multistability in the system. Because the FVS consists of nodes in every feedback in the network, removing the FVS creates an acyclic network. As demonstrated by Ren\'{e} Thomas in \cite{Thomas}, the cycles of a network are directly related to its multistability \cite{Richard}, so this acyclic network only has one attractor. Driving a node in a cycle to a specific state will remove the multistability of that cycle. Thus, overriding the state of every node in the FVS into their states in a specific attractor reduces the state space of the network so there is a single attractor, namely the desired one \cite{Mochizuki}. Controlling the entire FVS is a sufficient condition to drive the network into a given attractor, but it is not always necessary. Previous work has shown that control sets smaller than the FVS can drive the trajectory of a system into a desired attractor \cite{Mochizuki, Gates, Zanudo, Zanudo3}, which means that partial control of the FVS (e.g., control of a subset of the FVS) can be sufficient for attractor control when restricted to a specific attractor or to a particular model \cite{Zanudo5}. This is important because the size of the FVS can often be larger than the size of combinatorial interventions that can be implemented in biological experiments, which is usually restricted to 1-3 nodes. For example, the gene regulatory network in ascidian embryos studied by Kobayashi et al. in \cite{Kobayashi} has a FVS size of five, and it is near the limit of what can be currently controlled experimentally. Thus, it is of great interest to find FVS subsets that can drive a network to its natural attractors. At present, it is not known how to most effectively choose the correct FVS subsets, so in this work we aim to rank various FVS subsets based on their ability to drive the system into a desired attractor.

Although the FVS can be found without knowledge of the dynamics of the system, in this work we focus on Boolean dynamic models to evaluate the effectiveness of our methods. Boolean models characterize each node with two values, usually referred to as OFF (0) and ON (1), and with update equations usually based on logical operations. While Boolean models represent an approximation, they have been shown to capture the attractors and qualitative dynamics of various systems \cite{Abou,Wynn,Zanudo4}. Boolean models have a larger size (include more components) than continuous models. Larger networks will have larger FVSs, which will allow us to study cases in which a small fraction of the FVS is controlled. Furthermore, because the state space of Boolean models is finite, it is possible to comprehensively probe the effects of an intervention on the state space of the network. Thus, Boolean models provide a strong test bed for evaluating our methods.

The reduced T-LGL network in Fig. \ref{fig1}(a) was described with a Boolean model in \cite{Saadatpour}. The network's two fixed-point attractors -- apoptosis and survival -- are represented in Fig. \ref{fig1}(b), where white represents a node that is OFF and dark grey represents a node that is ON. We know controlling both of the minimal FVS nodes, S1P and FLIP, can drive the network into either of its attractors, but how does controlling only one of these two nodes (indicated by an dash-dotted outline) affect the dynamics? Controlling S1P alone can still drive the system to either of the two attractors [Fig. \ref{fig1}(c)]. However, controlling FLIP alone does not fix the state of any other node and both original attractors are still possible [Fig. \ref{fig1}(d)]. This difference between FVS members motivated us to search for an answer to the following questions: Are there network (topological) measures that differentiate S1P from FLIP and thus can predict that S1P outperforms FLIP in driving the system into an attractor? And in a general network, how can we identify which FVS subsets are most likely to control the network?

\section{\label{sec:KC}Key Concepts and Measures}
We evaluated the ability of seven topological measures to identify important FVS subsets. These measures capture a node's influence on the network, determined by either the local or global interactions of the node within the network. This influence tells us how much we expect the node to control the other nodes in the network. . The seven metrics are: out-degree, average inverse distance, PRINCE \cite{Vanunu,Santolini}, a modified version of PRINCE, CheiRank \cite{Zhirov}, involvement in the strongly connected component (SCC), and involvement in the network's positive cycles. These metrics fall into three categories: centrality measures, propagation measures, and cycle-based measures.

\begin{figure*}[ht]
\includegraphics[width=128mm]{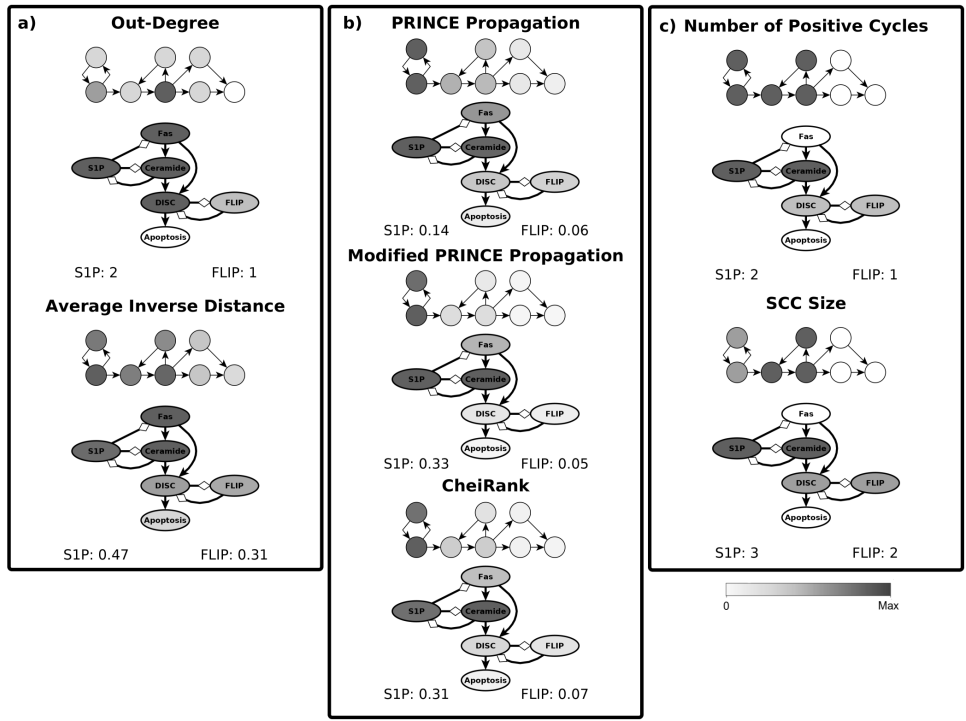}
\caption{\label{fig2}Illustration of the structure-based influence of the nodes in a network according to each of our seven topological metrics. These topological metrics are split into three categories: (a) centrality metrics, (b) propagation metrics, and (c) cycle-based metrics. Each metric is illustrated on a toy network and on the simplified T-LGL network. The shade of grey of each node is the node's conjectured influence over the other nodes in the network based on the given metric. For every metric, S1P is predicted to have a larger influence on the network than FLIP, which is consistent with the effect of these nodes on the attractors, illustrated in Fig. \ref{fig1}.}
\end{figure*}

The centrality measures -- out-degree and average inverse distance -- provide a baseline measure of how important each node is [Fig. \ref{fig2}(a)]. A node's out-degree provides a local measure of the node's influence on the network. The average inverse distance is a global measure of the node's influence, and quantifies how close the node is to every other node in the network. Both of these metrics are simple centrality measures, so they are expected to provide a baseline with which to gauge the predictive ability of the other metrics but may miss some of the intricacies of the network's dynamics.

The propagation measures -- PRINCE, a modified PRINCE, and CheiRank -- predict a node's influence on every other node in the network based on the steady state of a random walk algorithm [Fig. \ref{fig2}(b)]. PRINCE propagates a fixed perturbation on a target node through the network and attributes an “influence” value to every other node (see Methods). The propagation of influence over an edge is normalized by the out-degree of the source of the edge and by the in-degree of the target, so influence propagation is treated similarly to mass flow. Since the flow of information does not need to obey mass conservation, we propose and evaluate a modified PRINCE algorithm that only normalizes by out-degree (see Methods). For both measures we averaged the influence values to capture the perturbed node's influence over the entire network, with a higher average indicating a stronger influence (see Methods).

CheiRank works by conducting a random walk in a network in which all edges have the opposite direction, where at every step there is also a chance for the random walker to jump to a random node (see Methods). The influence assigned to a node by CheiRank is the probability that this node is the origin of this random walk. Thus, CheiRank is a measure of the “source-ness” of each node.

The cycle-based metrics aim to measure how much a subset contributes to the multistability of the network. To quantify this, the entire FVS is removed from the network to create an acyclic, mono-stable network. Then, the chosen subset and its corresponding edges are reintroduced to the network, and the number of positive cycles as well as the size of the largest SCC are measured [Fig. \ref{fig2}(c)]. More positive cycles presumes that the subset has larger control over sources of multistability in the network, and a larger SCC presumes that the node is more connected in the original network to nodes involved in multistability, so both of these metrics should approximately capture a node's level of control over the multistability in the network.

\begin{figure}
\includegraphics[width=\linewidth]{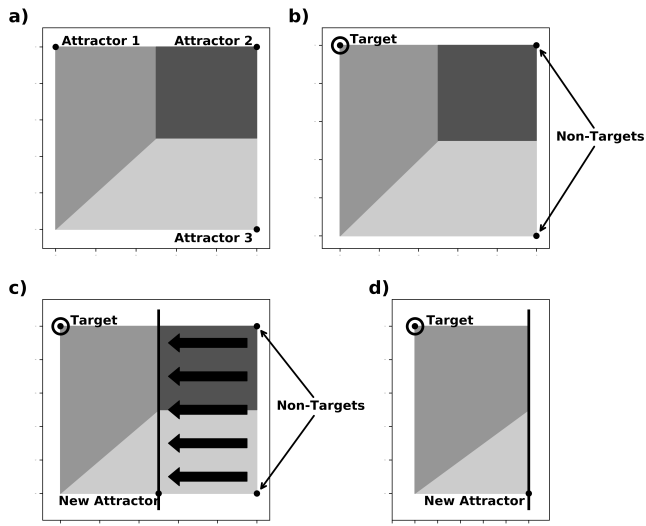}
\caption{\label{fig3}Conceptual figure demonstrating how interventions affect the attractors and their basins of attraction. (a) Representation of the state space of a network; the basins of attraction of the three attractors are indicated by different shades. (b) The three attractors and their basins are separated into two categories: Targets (attractor 1) and Non-targets (attractors 2 and 3). (c) An intervention is applied to the system, which decreases the state space. This restriction of the state space disallows both attractors 2 and 3 and creates a new attractor. (d) The new modified state space is used to determine \textit{To Control} and \textit{Away Control}. \textit{To Control} measures the increase in the size of the target basin while \textit{Away Control} measures the decrease in the size of the non-target basin.}
\end{figure}

When we introduce an intervention on a subset of the FVS, we measure the intervention's effect on the attractors and their basins to understand how well it controls the system. First, based on the intervention, we sort the attractors into targets and non-targets [Fig. \ref{fig3}(b)]. Applying the intervention to the system restricts the state space of the system [Fig. \ref{fig3}(c)], which changes the basins of attraction of the system's original attractors and may also introduce new attractor(s) [Fig. \ref{fig3}(d)]. When the goal is to drive the system into a target attractor, a successful intervention increases the basin of attraction of the target attractor and decreases the basins of the other attractors. We propose two metrics that measure these two aspects of control. \textit{To Control} is the measure of how well control of a FVS subset drives the network to the target attractor whereas \textit{Away Control} measures how well the same subset drives the network away from non-targeted attractors. The basis of both metrics is the normalized percent change in the relevant basin of attraction compared to the unperturbed system (see Methods). These two values may be equal but are not necessarily the same. For example, if an intervention decreases the basin of a non-target attractor while driving the system to a new attractor, it will not increase the basin of the target attractor, so the intervention will have a large \textit{Away Control} but a small \textit{To Control}.

As our aim is to quantify the effectiveness of the FVS subsets identified from each of our topological metrics, we perform a systematic evaluation of all interventions, i.e., for a FVS subset of size $L$ we consider all $2^L$ combinations of fixing each node in the ON or OFF state. For each of these $2^L$ interventions, we sort the attractors into target and non-target attractor(s). Each attractor wherein the states of every node are consistent with the intervention states becomes a member of the target attractors, while the other attractors are classified as non-target. Interventions consistent with every attractor are not informative, that is, an intervention that classifies every attractor as a target does not give any valuable information. Similarly, interventions that are not consistent with any attractors are only partially informative, that is, an intervention that classifies every attractor as a non-target can only give us information on how well the intervention can drive the system away from the attractors of the system (see Methods). We refer to interventions that classify some attractors as targets and others as non-target attractors (so both sets are non-empty) as fully informative interventions. For every fully informative intervention, we determine the values of \textit{To Control} and \textit{Away Control}; if the intervention is partially informative, we only determine the value of \textit{Away Control}.

For each FVS subset of size $L$, there are $2^L$ interventions, each with a value for \textit{To Control} and \textit{Away Control}. To summarize the values of all interventions associated with a FVS subset with a single quantity, we define an aggregate value of \textit{To Control} and \textit{Away Control} for a FVS subset as the maximal value over all its (partially) informative interventions. Once calculated, the value of these control metrics (control values) are compared with the topological metrics to determine how effective each metric is at identifying subsets that can control the network. We define a FVS subset as \textit{successful} if the aggregate value of the control metric of interest (\textit{To Control} or \textit{Away Control}) is larger than a success threshold, which we choose to be 0.9.

\begin{figure}[H]
\includegraphics[width=\linewidth]{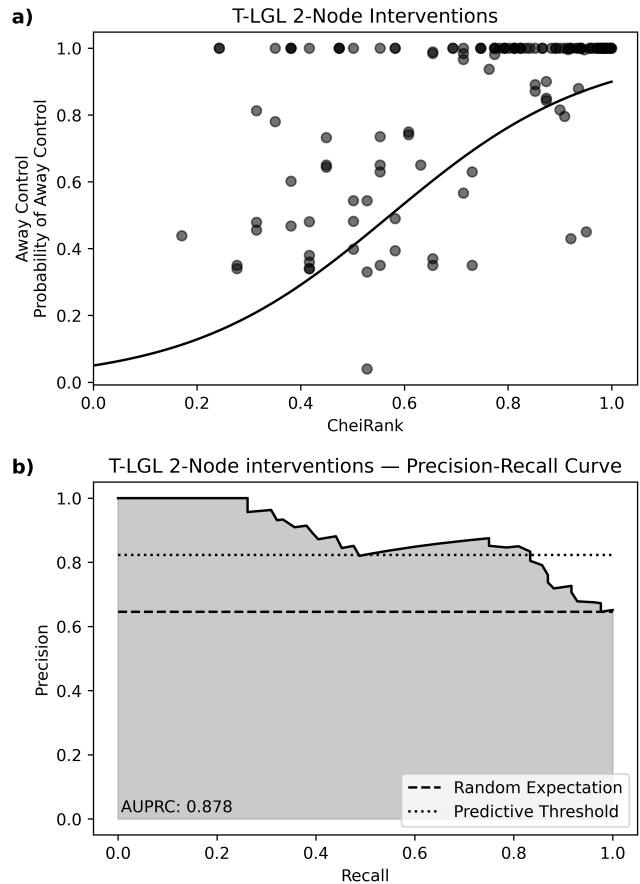}
\caption{\label{fig4} (a) Logistic regression and data points, and (b) precision-recall curve for \textit{Away Control} vs CheiRank obtained from driving all possible two-node FVS subsets in the T-LGL network. In panel (a), the x-axis indicates the value of the topological metric, the CheiRank in this case, for each FVS subset while the y-axis indicates the aggregate \textit{Away Control} value given by each FVS subset and the probability of being a successful FVS subset given by the logistic regression. To view this as a binary classification problem, we binarize our data by considering a FVS subset with an aggregate \textit{Away Control} value greater than 0.9 to be a \textit{successful FVS subset}. The AUPRC is the area under the precision-recall curve in panel (b), which is 0.878 for this data. The fraction of positive data points is 0.646, so the AUPRC predictive threshold (the AUPRC value above which we consider the topological metric effective at sorting) is 0.823. Because the AUPRC value of 0.878 is higher than this threshold, this indicates that the CheiRank is capable of sorting FVS subsets according to whether they are successful FVS subsets or not, and we say this AUPRC is predictive. Similar data was graphed for FVS subsets of various sizes (1 to 3) for each observed network, and the results of all of these graphs are displayed in Table \ref{tab1}.}
\end{figure}

We relate each of the seven topological metrics to the two control metrics using a logistic regression [Fig. \ref{fig4}(a)]. After creating the logistic regression, the sorting strength of each topological metric when classifying successful FVS subsets is determined based on the area under the precision-recall curve (AUPRC) metric (see Methods). We determined the precision ($\frac{TruePositives}{TruePositives+FalsePositives}$) and recall ($\frac{TruePositives}{TruePositives+FalseNegatives}$) of the logistic regression by comparing the successful FVS subsets, based on the control metric of interest, to the predicted successful FVS subsets based on setting a threshold on the regressor variable (the x-axis value, i.e., topological metric value).

The precision and recall values for each value of the regression variable (topological metric) are then plotted against each other; the area under this curve is the AUPRC [Fig. \ref{fig4}(b)]. If the AUPRC value is equal to the fraction of positive data points, the topological metric cannot sort the FVS subsets better than randomly sorting subsets. We define the AUPRC predictive threshold as the AUPRC value that is greater than halfway between the fraction of positive data points and the maximum AUPRC value of 1. Thus, a regressor variable is predictive if it sorts the binarized values of a control metric with an AUPRC value above the AUPRC predictive threshold. For example, if the fraction of positive data points is 0.75, the AUPRC predictive threshold is 0.875, so an AUPRC greater than 0.875 would be considered predictive. Calculating these AUPRC values for each of our topological metrics across different networks and subset sizes allows us to determine how well they can identify successful FVS subsets, or in other words, how well they predict a FVS subset's ability to drive the system into the target attractor(s) or away from the non-target attractor(s).

\section{\label{sec:Results}Results}
We investigated an array of four different networks, corresponding to well-established Boolean models of biological systems, namely the T-LGL network \cite{Zhang}, NSCLC network \cite{Udyavar}, an FA/BRCA variant network \cite{Rodriguez}, and a Helper T Cell Differentiation network \cite{Naldi}. For these networks, we need to identify their minimal FVS, which are not necessarily unique, but identifying a minimal FVS is an NP-hard problem. Instead, and following \cite{Zanudo}, we identify near-minimal FVSs using one of the efficient methods available (the simulated annealing algorithm introduced in \cite{Galinier}). We analyzed all FVS subsets of one to three nodes from multiple near-minimal FVSs we identified. The size of these networks range from 28 to 60 nodes, and the size of the near-minimal FVSs range from 8 to 17 nodes (see Table S4).

\subsection{\label{subsec:TopoRes}Propagation metrics are the best at classifying successful FVS control subsets}
To characterize the predictive power of each of the seven topological metrics, we determined their associated AUPRC values for each network and FVS subset size and used these AUPRC values to determine the predictive power of each topological metric in the task of classifying successful control subsets. We summarize these results in Table \ref{tab1}.

The top of Table \ref{tab1} shows the percentage of cases (out of 12 total cases, where each case corresponds to a single FVS subset size of one of the networks) in which the indicated metric was predictive according to the AUPRC. Propagation metrics had the highest \% of cases of all seven metrics for both \textit{To Control} and \textit{Away Control} and cycle-based metrics had the lowest \% of cases for both control metrics. The bottom of Table \ref{tab1} shows the rank of the AUPRC values for each of the seven metrics averaged over all 12 cases, where a lower rank corresponds to a higher AUPRC value. Consistent with what we found with the \% of cases, propagation metrics had the lowest average rank of all seven metrics for both control metrics. Although the order of centrality and cycle-based metrics was not consistent for both control metrics, cycle-based metrics had the highest average rank of all metrics for the \textit{To Control} metric. In summary, we found that propagation metrics had the highest predictive power out of all metrics (highest \% of cases in which they were predictive and lowest average AUPRC rank) and that cycle-based metrics had the lowest predictive power out of all metrics (lowest \% of cases in which they were predictive and highest average AUPRC rank for the \textit{To Control} metric.).

\begin{table*}
\caption{\label{tab1}Table summarizing results of the AUPRCs for each topological metric measured across 4 networks with all FVS subsets of size 1, 2, or 3. We measured the percentage of cases where the AUPRC value was greater than the AUPRC predictive threshold (top) and the average rank of the AUPRC value (a lower rank corresponds to a higher AUPRC value) among the seven metrics (bottom). Propagation metrics sort FVS subsets the best when compared to the other topological metrics, while the cycle-based metrics sort FVS subsets the worst.}
\begin{ruledtabular}
\begin{tabular}{l*{7}{c}}
&\multicolumn{2}{c}{Centrality}&\multicolumn{3}{c}{Propagation}&\multicolumn{2}{c}{Cycle-based}\\
\cline{2-3}\cline{4-6}\cline{7-8}
&Out-degree&AverageDistance&PRINCE&Modified PRINCE&CheiRank&Positive Cycles&SCC Size\\
\hline
\multicolumn{8}{c}{\% of cases where AUPRC is above the predictive threshold}\\
\hline
\textit{To Control}&50.00\%&41.67\%&58.33\%&66.67\%&58.33\%&25.00\%&41.67\%\\
\textit{Away Control}&66.67\%&58.33\%&66.67\%&83.33\%&83.33\%&58.33\%&41.67\%\\
\hline
\multicolumn{8}{c}{Average rank of AUPRC values}\\
\hline
\textit{To Control}&4.58&4.29&2.83&2.92&3.50&4.33&4.38\\
\textit{Away Control}&3.33&4.38&3.08&2.46&2.71&5.04&6.04
\end{tabular}
\end{ruledtabular}
\end{table*}

\subsection{\label{subsec:InterRes}Intersections of top-ranking FVS subsets for each topological metric can improve FVS subset identification}

\begin{figure}
\includegraphics[width=\linewidth]{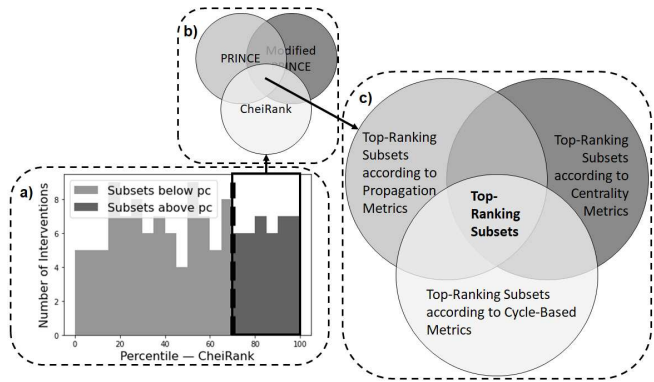}
\caption{\label{fig5}Intersecting the sets containing the highest-valued FVS subsets of distinct topological metrics identifies FVS subsets hypothesized to have a large level of control over the network dynamics. (a) Based on a chosen cutoff percentile, we identified the highest-valued subsets in a distribution of the topological metric. Here, the 70th percentile cutoff (pc) gives us the top 30\% of subsets according to the CheiRank. (b) These top-ranking CheiRank subsets were then intersected with the subsets in the top 30\% of the other two propagation metrics (PRINCE and Modified PRINCE) to identify the FVS subsets that perform well for all propagation metrics. (c) The intersection set of the propagation metrics was intersected with the intersection set of the other metrics to identify the FVS subsets that perform well for every topological metric. Intersection sets contain FVS subsets that are hypothesized to better control the network, so they are expected to contain a lower percentage of false positives than each topological metric.}
\end{figure}

\begin{figure*}
\includegraphics[width=\textwidth]{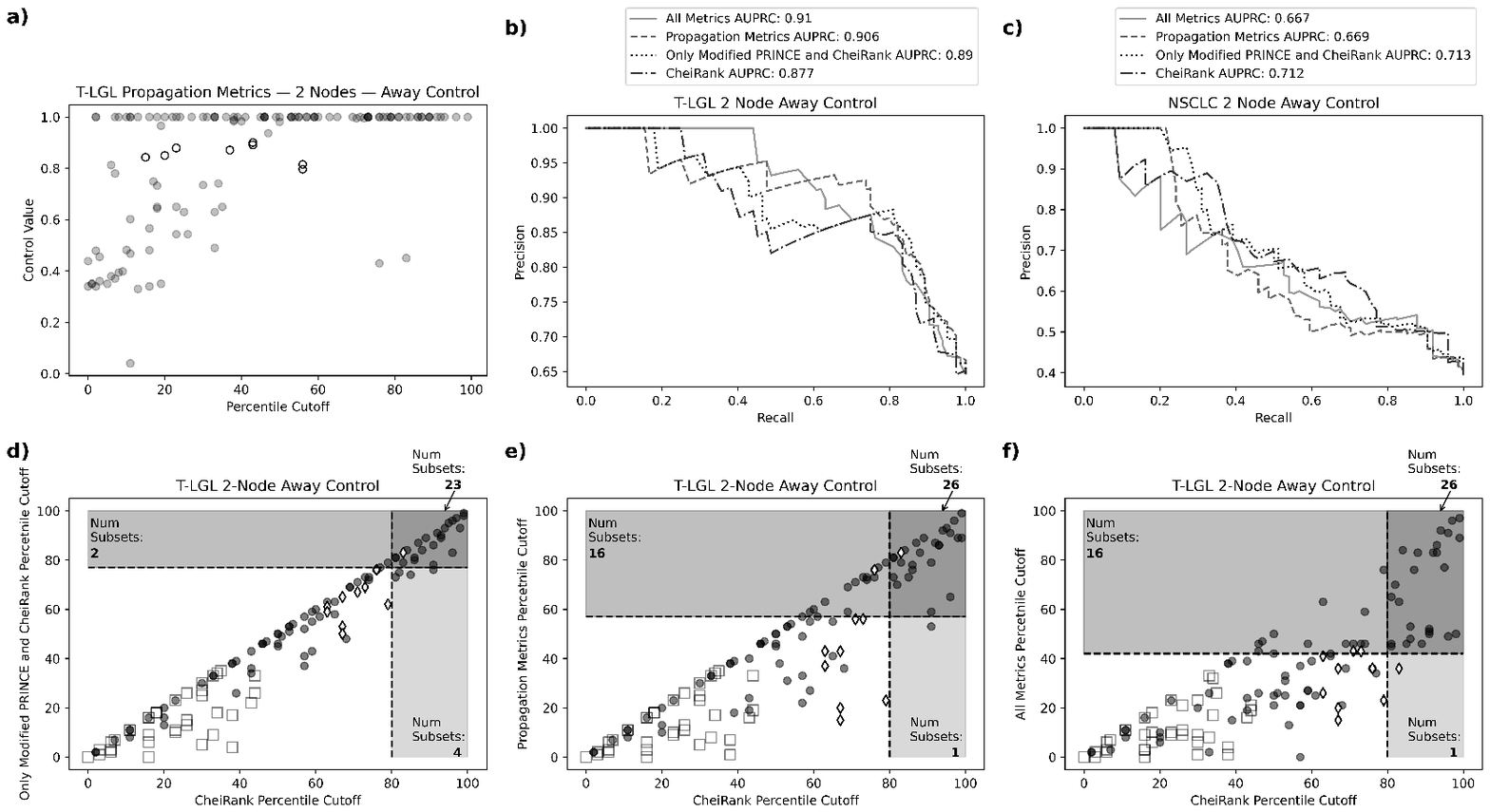}
\caption{\label{fig6} Intersection metrics improve the number of FVS subsets identified in the high precision and low recall regime of the system. (a) Scatter plot showing the propagation intersection metric percentile cutoff value and the \textit{Away Control} value of FVS subsets for the T-LGL two-node intervention case. Each FVS subset's value on the x-axis represents the maximum percentile cutoff value at which the FVS subset appears in the propagation intersection metric. (b-c) Precision-recall curves of CheiRank and the intersection metrics for \textit{Away Control} in the two-node intervention case of the T-LGL network (panel b) and NSCLC networks (panel c). (d-f) Percentile cutoff value of CheiRank vs percentile cutoff value of each of the studied intersection metrics. Filled black circles denote successful FVS subsets (based on the binarized value of the control metric) and open shapes denote unsuccessful FVS subsets. Among the open shapes, diamonds indicate FVS subsets that have a high percentile cutoff (\textgreater 50\%) for CheiRank. On each plot, we indicate the percentile cutoff where the precision drops below 0.95 for CheiRank (vertical dotted line) and for the intersection metric (horizontal dotted line). The top right quadrant contains the FVS subsets identified by both CheiRank and the intersection metric; the top left quadrant contains the FVS subsets identified by the intersection metric but not CheiRank; the bottom right quadrant contains the FVS subsets identified by CheiRank but not the intersection metric. As more metrics are included in an intersection metric, the percentile cutoff of the unsuccessful FVS subsets decreases farther than the percentile cutoff of successful FVS subsets. This shifts the 0.95 precision point of the intersection metric (horizontal line) lower on each successive panel, below the percentile cutoff of successful FVS subsets, which shift only by a small amount. Because more FVS subsets are above the 0.95 precision point in the intersection metrics than in CheiRank, the top left quadrant of subsequent panels contains more FVS subsets than the first panel. Using intersection metrics improves the precision of the top ranking subsets, but including too many metrics in the intersection can make the intersection metric too strict and can decrease or not benefit its predictive power. This is illustrated by the equal number of identified FVS subsets between the propagation intersection metric and the all metrics intersection metric.}
\end{figure*}

The number of FVS subsets that have a high value in a topological metric but a low control value reflects the metric's inability to fully capture the dynamics of the system. We hypothesized that intersections of the FVS subsets predicted to successfully control the network by each topological metric would reduce the number of FVS subsets that have a high value in a topological metric but a low control value. To do this intersection, we set a fixed percentile cutoff, and for each topological metric, we identified the set of FVS subsets that was above the percentile cutoff (see Methods and Fig. \ref{fig5}). The sets of above-cutoff FVS subsets [box in Fig. \ref{fig5}(a)] for each topological metric of interest were then intersected with each other resulting in a set of FVS subsets that are hypothesized to better control the network and contain a lower percentage of false positives than each metric individually. We use the term intersection metrics to refer to the metrics defined by intersecting topological metrics, and we denote them according to the metrics that are included in the intersection. In an intersection metric, we assign to each FVS subset the highest percentile cutoff value at which the FVS subset appears in the intersected topological metrics (see Methods). To evaluate how well an intersection metric performs as more metrics are included in the intersection and to determine which combinations of the most predictive metrics best approximate FVS subsets' level of control over the network dynamics, we tested three different intersection metrics: 1) an intersection of all seven metrics, 2) an intersection of just the propagation metrics, and 3) an intersection of the Modified PRINCE and CheiRank metrics. As an example, Fig. \ref{fig6}(a) shows the percentile cutoff value and \textit{Away Control} value for the FVS subsets of the intersection metric that includes the three propagation metrics.

To analyze each intersection metric, we took a similar approach to what we did in Section \ref{subsec:TopoRes}. We obtained the AUPRC values for each intersection metric, the percentage of AUPRC values above the predictive threshold (Table \ref{tab2} top), and the average rank of the AUPRC values among the three selected intersection metrics and all seven topological metrics (Table \ref{tab2} bottom). The intersection metrics are predictive (have an AUPRC above the predictive threshold) in more cases than the non-propagation metrics, as reflected by their higher percentage of cases (see Table \ref{tab1} for the non-propagation metrics and Table \ref{tab2} for the intersection metrics). The intersection metrics always have an average rank that is lower than that of the non-propagation metrics (Table S1), and they perform similarly to the propagation metrics (Table \ref{tab2}), which performed the best among all individual topological metrics (Table \ref{tab1}). Among the intersection metrics, the Modified PRINCE and CheiRank intersection metric is predictive in the highest percentage of cases (75\% in both \textit{To Control} and \textit{Away Control}), it has the lowest average rank for \textit{Away Control}, and it has the second lowest average rank for \textit{To Control}. Notably, and contrary to our expectations, the intersection metrics do not outperform the Modified PRINCE metric, which has the first or second highest percentage of cases in which it is predictive and the lowest average rank of AUPRC values among all metrics (Table \ref{tab2}). 

\begin{table*}
\caption{\label{tab2} Table summarizing the results of the AUPRCs for each intersection metric. The percentage of cases where the AUPRC is above the AUPRC predictive threshold (top) and the average rank of the intersection metrics (bottom) are displayed. These AUPRC values are ranked among the three intersection metrics and the seven topological metrics. The percentage of cases for each intersection metric and their rank are similar to those of the individual propagation metrics. Note that the ranks range from 1-10 in this table, but they range from 1-7 in the bottom of Table \ref{tab1}, so the ranks between the tables cannot be directly compared. The ranks for the three propagation topological metrics are also shown in the table as they were the best performing individual topological metrics. }
\begin{ruledtabular}
\begin{tabular}{l*{6}{c}}
&PRINCE&Modified PRINCE&CheiRank&Modified PRINCE and CheiRank&Propagation&All\\
\hline
\multicolumn{5}{c}{Percentage of cases where AUPRC is above Success Cutoff}\\
\hline
\textit{To Control}&58.33\%&66.67\%&58.33\%&75.00\%&58.33\%&58.33\%\\
\textit{Away Control}&66.67\%&83.33\%&83.33\%&75.00\%&66.67\%&75.00\%\\
\hline
\multicolumn{5}{c}{Average rank of AUPRC values}\\
\hline
\textit{To Control}&5.71&4.04&4.83&4.67&5.83&4.38\\
\textit{Away Control}&5.71&3.42&3.88&3.71&5.21&4.75
\end{tabular}
\end{ruledtabular}
\end{table*}

Although the Modified PRINCE and CheiRank intersection metric performed the best among the intersection metrics and the propagation intersection metric performed the worst, we noticed that relying on the AUPRC to rank metrics can mask a property that is desirable in our setting: the ability of a topological metric to be predictive in the high precision and low recall regime. This regime is important because it is where we expect topological metrics to excel. For each FVS subset we expect that having a top rank in a predictive topological metric is sufficient but not necessary to have a high control metric value; this corresponds to the high precision and low recall regime. Note that the reason we do not necessarily expect topological metrics to fully capture the control metric values is that the control values depend on the dynamic information in the model and not solely on the network topology.

For example, in the two-node, \textit{Away Control} case for the NSCLC network, the AUPRC for the propagation intersection metric is 0.669, which is below the AUPRC predictive threshold of 0.698, but on the precision-recall curve [Fig. \ref{fig6}(c)], the propagation intersection metric has the highest recall before the precision drops from 1.0, i.e., the propagation intersection metric identifies the most successful \textit{Away Control} FVS subsets before encountering an unsuccessful \textit{Away Control} FVS subset. Despite performing the worst according to the AUPRC values, the propagation intersection metric performs the best among all studied metrics (has the lowest average rank, 2.83) when comparing metrics according to the number of the successful FVS subsets they identify before the precision drops below 0.95 (Supplemental Material, Table S3). The propagation intersection metric outperforms Modified PRINCE, which performed the second best (2.96), and the Modified PRINCE and CheiRank intersection metric, which performed the third best (3.33). Thus, while the AUPRC is a good indicator of how well every subset is sorted based on its binarized control value, it can fail to capture how well a metric performs in the high precision and low recall section of the precision-recall curve, as with the propagation intersection metric in this case. To focus on this regime of the precision-recall curve, we find the percentile cutoff value for which the precision drops below 0.95, and use this 0.95 precision percentile cutoff value, which we refer to as the 0.95 precision point, to identify the high precision and low recall FVS subsets. 

To illustrate how the intersection metrics perform in identifying successful FVS subsets when compared to other intersection metrics, we plot the percentile cutoffs of each intersection metric against the percentile cutoffs of the CheiRank in Fig. \ref{fig6}(d-f). Successful FVS subsets are represented by filled circles and unsuccessful FVS subsets by open shapes. We use horizontal and vertical dotted lines to mark the 0.95 precision point for CheiRank and the intersection metric respectively. These two lines split each figure panel into four quadrants, which we can use to evaluate how the FVS subsets identified by the intersection metric differ from those of CheiRank. The top right quadrant corresponds to the FVS subsets identified by both CheiRank and the intersection metric, the bottom right quadrant corresponds to the FVS subsets that CheiRank identifies but are absent from the intersection metric, and the top left quadrant corresponds to the new FVS subsets that the intersection metric identifies that were not identified with CheiRank alone. As shown in Fig. \ref{fig6} (d-f), the number of FVS subsets with high percentile cutoff value and a binarized control value of one (control value \textgreater 0.9) increases as more metrics are added to the intersection (25 for the Modified PRINCE and CheiRank intersection metric to 42 for the propagation intersection metric). The reason for this is that the unsuccessful FVS subsets (open data points) have a larger decrease in percentile cutoff (shift down farther in the plot) than the successful FVS subsets (filled data points), so more FVS subsets are in the top section before the precision drops below 0.95. Thus, as we increase the number of metrics in our intersection metric, high percentile cutoff and high control FVS subsets remain above the 0.95 precision point value (horizontal line) while high-percentile cutoff and low-control FVS subsets shift to a lower percentile cutoff value, which lowers the 0.95 precision point cutoff value and increases the number of FVS subsets above cutoff, creating a higher recall before the precision decreases.

As an example of the improvement brought by intersection metrics, consider the cluster of high CheiRank percentile cutoff and low control value FVS subsets (open diamonds). As more metrics are added to the intersection in each subsequent panel, these FVS subsets' percentile cutoffs decrease (i.e., they are shifted down on the y-axis) farther than the FVS subsets with a similar percentile cutoff value and successful control value. Most of the FVS subsets marked with open diamonds (8 out of 10) contain the node JAK, which is part of a 2-node negative feedback loop. Negative feedback loops do not contribute to the multistability of the network, so their control has a weak contribution to driving the system into or away from an attractor. Since CheiRank does not consider edge signs, it assigns JAK the 4th highest value among all nodes. JAK ranks 15th for PRINCE, which does consider edge signs, so in Fig. \ref{fig6}c, we see these open diamonds have much lower percentile cutoff values. As more topological metrics are added to the intersection metric, every FVS subset's percentile cutoff will decrease or stay the same. In particular, when a FVS subset is bottom ranking for a newly introduced topological metric, the decrease in its percentile cutoff will be larger than that of subsets that are top ranking in the new metric. This also occurs for the other high percentile cutoff FVS subsets that don't include JAK (2 out of 10). These subsets always contain either CTLA4 or TCR, which form an isolated negative cycle, so when the positive cycles metric is included in the intersection, these FVS subsets also greatly decrease their percentile cutoff.

Based on the above results, intersection metrics appear to be more predictive than the individual metrics in the high precision and low recall regime, and we attribute this to the ability of the metrics in the intersection to complement each other's weakness. We further support these results with Tables S2 and S3 in the Supporting Material where we indicate the 0.95 precision point for all four networks (Table S2) and the number of FVS subsets above the 0.95 precision point (Table S3) for subsets of size 1 to 3 using all three intersection metrics and CheiRank. Overall, this data indicates that the percentile cutoff decreases and more FVS subsets are identified as more metrics are added to the intersection metric. However, it is possible for the intersection to become too strict if one of the metrics in the intersection sorts the data poorly. This negatively affects the number of subsets identified, so when using the all metric intersection fewer FVS subsets are identified than in the propagation intersection. These results show that the intersection metrics have higher predictive power than any individual topological metric in the high precision and low recall regime.

\begin{table}
\caption{\label{tab3}The number and precision of the two-node FVS subsets identified in intersections on the observed networks. The intersection percentiles were determined using the maximum (93 for \textit{To Control} and 90 for \textit{Away Control}), and third quartile (83 for \textit{To Control} and 72.75 for \textit{Away Control}) of the recorded propagation intersection metric 0.9 precision crossing points. Subsets were identified using the propagation intersection metrics at both percentile cutoffs. The maximum percentile cutoff does normally have better precision than the third quartile, but less subsets are identified. Furthermore, when comparing our results to the subsets predicted by a balanced and unbalanced logistic regression on the control value versus percentile cutoff data, we see that both logistic regressions find more FVS subsets but are less accurate.}
\begin{ruledtabular}
\begin{tabular}{m{2cm} c r c r c r c r}
\multirow{2}{2cm}{Percentile\\Cutoff}&\multicolumn{2}{c}{T-LGL}&\multicolumn{2}{c}{NSCLC}&\multicolumn{2}{c}{\shortstack{FA/BRCA\\Var \#1}}&\multicolumn{2}{c}{\shortstack{Helper \\T Cells}}\\
\cline{2-3}\cline{4-5}\cline{6-7}\cline{8-9}
&Size&\multicolumn{1}{c}{Acc}&Size&\multicolumn{1}{c}{Acc}&Size&\multicolumn{1}{c}{Acc}&Size&\multicolumn{1}{c}{Acc}\\
\hline
&\multicolumn{8}{c}{\textit{To Control}}\\
\cline{2-9}
Max (93)&2&100\%&5&100\%&2&0\%&3&100\%\\
Q3 (83)&8&62\%&16&94\%&3&0\%&7&100\%\\
LogReg\newline(Balanced)&18&78\%&84&45\%&14&21\%&26&92\%\\
LogReg\newline (Unbalanced)&16&75\%&38&63\%&0&N/A&53&81\%\\
\noalign{\smallskip}
&\multicolumn{8}{c}{\textit{Away Control}}\\
\cline{2-9}
Max (90)&5&100\%&7&100\%&3&100\%&4&100\%\\
Q3 (72.75)&30&93\%&27&78\%&6&100\%&15&93\%\\
LogReg\newline(Balanced)&70&90\%&86&51\%&20&90\%&32&84\%\\
LogReg\newline (Unbalanced)&82&84\%&57&60\%&27&85\%&63&75\%\\
\end{tabular}
\end{ruledtabular}
\end{table}

\subsection{\label{subsec:PCRes}Development and testing of a generalizable percentile cutoff based on our Boolean networks}

\begin{figure}
\includegraphics[width=\linewidth]{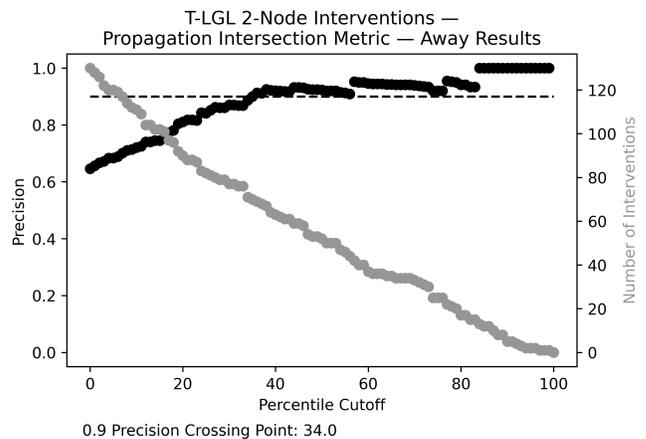}
\caption{\label{fig7}Example graph of the precision and number of interventions identified for the all metric intersection metric percentile cutoffs. The two-node, \textit{Away Control} case of the T-LGL network is shown. Black dots represent the precision of the FVS subsets identified at each percentile cutoff, and grey dots represent the number of FVS subsets that are identified. At a percentile cutoff of 0, every FVS subset is included, precision is at a minimum, and the number of subsets is at a maximum. As the percentile cutoff increases, the precision increases as subsets are filtered out. Eventually, the precision reaches 1.0, indicating that all the unsuccessful subsets are filtered out. Once the percentile cutoff gets too restrictive, no FVS subsets are identified, so the precision drops to 0. The dotted line indicates a precision of 0.9, which is used to determine the 0.9 precision crossing point.}
\end{figure}

Knowing that at certain percentile cutoffs the intersection metrics can better identify successful subsets than individual metrics, we used our four networks to determine which percentile cutoff was the most appropriate for picking successful FVS subsets. We aimed to find a percentile cutoff with high precision while maximizing the number of identified FVS subsets. This involves a trade-off because as the precision increases the number of FVS subsets decreases, as shown in Fig. \ref{fig7}. For the rest of the manuscript, we chose the point where the precision first crosses 0.9 from below (dashed line on Fig. \ref{fig7}), which we refer to as the 0.9 precision crossing point. These crossing points were identified for all four networks on FVS subsets of size 1, 2, and 3, for the propagation intersection metric, separately for \textit{To Control} and \textit{Away Control}.

Based on the full set of these propagation intersection metric 0.9 precision crossing points for each control measure, two percentile cutoff values were chosen, which we expect will be able to identify successful FVS subsets when applied to other Boolean networks. The maximum of the 0.9 precision crossing points indicates the value that is most likely to achieve a high precision, but it is possible that it does not detect any FVS subsets. To increase the likelihood of identifying a FVS subset, we also evaluated the third quartile percentile cutoff because it is less stringent, but is still expected to have high precision. For \textit{To Control}, we found that the 93 percentile was the maximum cutoff value and the 83 percentile was the third quartile. For \textit{Away Control}, these two values were the 90 and 72.75 percentile cutoffs respectively. The \textit{Away Control} values are lower because it is easier for a subset to achieve high \textit{Away Control} than \textit{To Control}, so a less stringent percentile cutoff performs better.

To verify that our method does identify a sufficient number of successful FVS subsets with high precision, we reapplied the percentile cutoffs identified by our method to the four networks used to derive these percentile cutoffs. For each control measure, both cutoffs were applied on the propagation intersection metric (Table \ref{tab3}) and the all metric intersection metric (Table S5). In Table \ref{tab3}, the number of identified FVS subsets of size 2 with percentile cutoff higher than these values was recorded as well as the precision in terms of being successful FVS subsets; data for FVS subsets of size 1 or 3 are presented in Table S5. In Table \ref{tab3}, the FVS subsets identified using the maximum percentile cutoff value have a precision of 100\% in all cases except for the \textit{To Control} case of the FA/BRCA variant network, and have a precision higher than 85\%/95\% in the \textit{To Control}/\textit{Away Control} cases respectively (Table S5). As expected, the third quartile identifies more subsets and has a lower precision than the maximum, but overall using the third quartile percentiles still identifies FVS subsets that achieve a precision above 60\% for \textit{To Control} and a precision above 75\% for \textit{Away Control}. As a point of reference, we also show the size and precision of the identified FVS subsets for both a balanced and unbalanced logistic regression (see Methods). When compared to the logistic regressions, finding the FVS subsets above these percentile cutoffs is more stringent but also typically more precise because the logistic regression does not focus on the high precision and low recall regime.

While using these percentile cutoffs has a higher precision than the logistic regressions in Table \ref{tab3}, this is not true in the \textit{To Control} case of this variant of the FA/BRCA network. This outlier results from the node ICL being crucial for driving the system to the correct attractor, but of all the single nodes whose interventions are fully informative, ICL has the lowest topological values. Thus, if a two- or three-node subset has a high percentile cutoff value, it likely won't contain ICL and won't drive the system to the correct attractor, so the subsets with percentile cutoffs above our percentile cutoff values won't be successful. Logistic regression is still able to identify some successful FVS subsets because the logistic fit leverages the observation that lower percentile cutoffs tend to have higher control values for this model. This observation cannot be incorporated in our percentile cutoff method, which assumes that FVS subsets with high percentile cutoffs are correlated with high control values. This outlier demonstrates that there are complexities of the dynamics of a network that may not be easily captured by topological metrics. Nevertheless, because applying the percentile cutoff values to identify FVS subsets on our networks was able to identify a sufficient number of FVS subsets that have a precision higher than 85\% when using the maximum percentile cutoff, we hypothesize that the percentile cutoff values identified by our method would have a similar performance when applied to other networks.

To test this hypothesis, we applied the percentile cutoffs identified by our method to identify FVS subsets that are expected to have high control values in a second array of Boolean models of biological networks. These networks were: a geroconversion network \cite{Verlingue}, two more variants of the FA/BRCA network, and two variants of a MAPK network \cite{Grieco} (see Table S4). We identified and found the precision of the subsets using our maximum and third quartile percentile cutoff values (see Table S6). Table \ref{tab4} shows the results of the propagation intersection metric for two-node FVS subsets. Many of the top-ranking FVS subsets had a binarized control value of 1 on their respective networks. However, this second array of network models had fewer nodes than our original array, so the most restrictive choice often didn't result in any FVS subsets being identified (see Table S6). So when applying this method to another network, it may be useful to start with the third quartile percentile values (83 for \textit{To Control} and 72.75 for \textit{Away Control}) despite their potentially lower precision. The results of Table S6 show that the sets of FVS subsets with percentile cutoffs above the third quartile percentile cutoff values are of sufficient size and high precision on other networks too. This shows that our percentile cutoff method is able to identify FVS subsets in the high precision and low recall regime using only the structure of the network.

\begin{figure}[h]
\includegraphics[width=\linewidth]{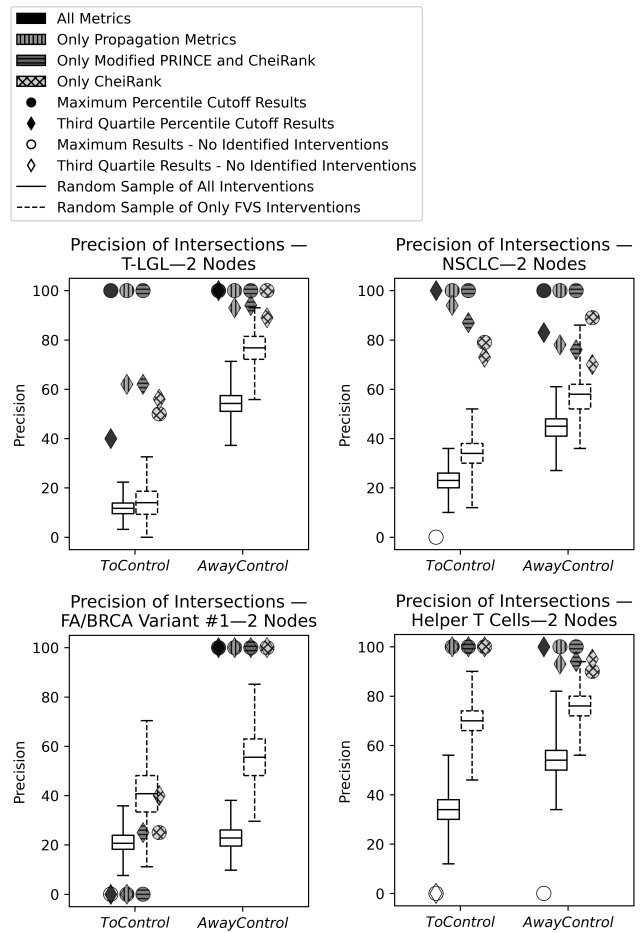}
\caption{\label{fig8}Plots of the precision of identifying successful FVS subsets on the original array of networks using our percentile cutoff method and box-and-whisker plots created by bootstraps on random samples of node subsets. There are two random samples, one is generated from a random sample of every possible subset and the second is generated from a random sample of only FVS subsets. Each graph shows \textit{To Control} on the left and \textit{Away Control} on the right. The graphs use four different shades of grey and patterns to differentiate between the intersection metrics and use two different shapes to differentiate between which percentile cutoff was used to generate the precision. These results indicate that when node subsets are identified using FVS subsets and our percentile cutoff method, the precision of these FVS subsets is higher than the precision of a set of randomly chosen FVS subsets and much higher than the precision of a set of randomly chosen node sets.}
\end{figure}

\begin{figure}[h]
\includegraphics[width=\linewidth]{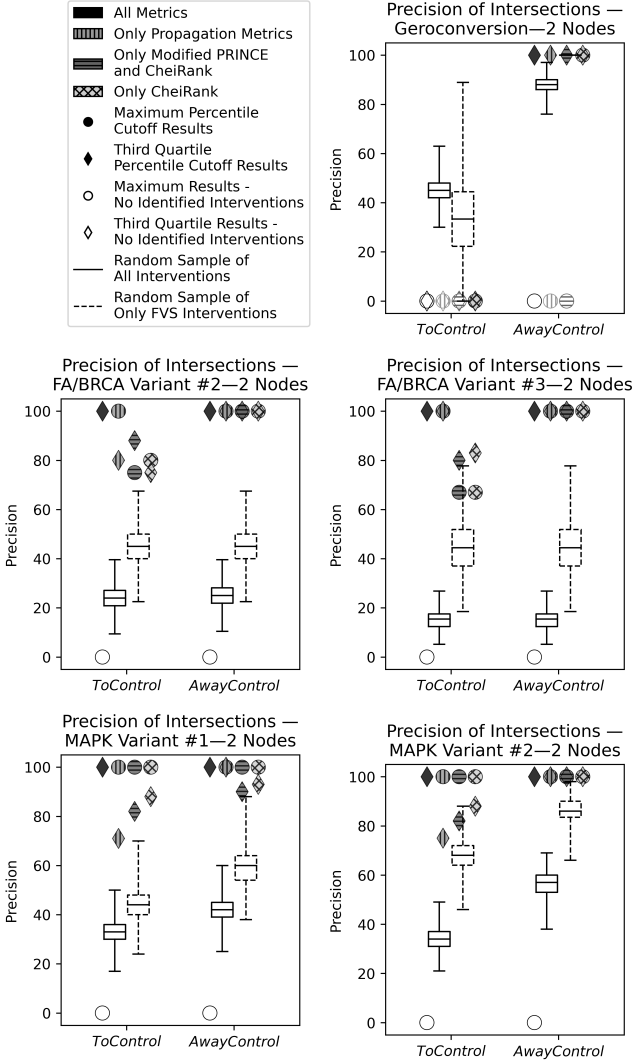}
\caption{\label{fig9}Plots of the precision of identifying successful FVS subsets on a second array of networks using our percentile cutoff method and box-and-whisker plots created by bootstraps on random samples of subsets. These results are similar to Fig. \ref{fig8}, indicating that using only the structure of a network and the percentile cutoffs identified by our method, the identified FVS subsets typically have a higher precision than a set of randomly chosen nodes or FVS subsets.}
\end{figure}

\begin{table*}
\caption{\label{tab4} Precision of the FVS subsets identified using the previously identified percentile cutoffs the new array of test networks. The third quartile values (83 for \textit{To Control} and 72.75 for \textit{Away Control}) are used to guarantee FVS subsets are identified on these networks because they have fewer nodes than our original set of networks. Using the propagation intersection metric, the number and precision of the identified two node FVS subsets was found. On these new networks, many of the predicted FVS subsets were successful in terms of their \textit{To Control} values and every predicted FVS subset was successful in terms of their \textit{Away Control} values, so these percentile cutoff values are expected to be generally applicable.}
\begin{ruledtabular}
\begin{tabular}{m{2cm} c r c r c r c r c r}
\multirow{2}{2cm}{Percentile\\Cutoff}&\multicolumn{2}{c}{Geroconversion}&\multicolumn{2}{c}{\shortstack{FA/BRCA\\Var \#2}}&\multicolumn{2}{c}{\shortstack{FA/BRCA\\Var \#3}}&\multicolumn{2}{c}{\shortstack{MAPK\\Var \#1}}&\multicolumn{2}{c}{\shortstack{MAPK\\Var \#2}}\\
\cline{2-3}\cline{4-5}\cline{6-7}\cline{8-9}\cline{10-11}
&Size&\multicolumn{1}{c}{Acc}&Size&\multicolumn{1}{c}{Acc}&Size&\multicolumn{1}{c}{Acc}&Size&\multicolumn{1}{c}{Acc}&Size&\multicolumn{1}{c}{Acc}\\
\hline
\textit{To Control} (83)&0&N/A&5&80\%&4&100\%&7&71\%&8&75\%\\
\textit{Away Control} (72.75)&1&100\%&11&100\%&7&100\%&15&100\%&14&100\%\\
\end{tabular}
\end{ruledtabular}
\end{table*}

\subsection{\label{subsec:BoxAndWhisk}The identified FVS subsets significantly outperform random samples of node subsets}
To confirm that this method identifies successful FVS subsets with high precision, we compared our results to random samples of node subsets. Two different random samples were generated: a random sample of every possible subset and a random sample of FVS subsets. To approximate the precision value of all subsets, we use bootstrapping in each random sample (see Methods).Twenty-five randomly selected single node subsets were chosen, or every individual FVS node was chosen if there were less than 25 total options. For 2 and 3 node subsets, 100 random subsets were picked when looking at all possible subsets and 50 random FVS subsets were picked when only looking at the FVS subsets. After a random sample was chosen, each subset's $2^L$ possible interventions were simulated to determine their \textit{To Control} and \textit{Away Control} values. We use the random subsets' binarized control values to determine their precision and compare it to the FVS subsets identified using the intersection metrics.

The distribution of precision values from bootstrapping is plotted as a box-and-whisker plot together with the precisions obtained when using the maximum percentile cutoff and third quartile percentile cutoff determined in the previous analysis of the FVS subsets, shown in Fig. \ref{fig8} for the original array of networks and Fig. \ref{fig9} for the new array of networks. In our networks, the precisions of the FVS subsets identified using our method significantly outperformed the random samples' precisions. This is true for each intersection metric, and even works partially well for CheiRank. The only exceptions were the \textit{To Control} cases of both the first variant of the FA/BRCA network and geroconversion network cases and when the intersection didn't identify any subsets (indicated by an open shape).

\subsection{\label{subsec:NetworkRes}Network specific results}
In the following, we analyze the FVS subsets identified using our percentile cutoff method and discuss their biological significance for two of the analyzed networks.

\subsubsection{\label{subsubsec:TLGL}T-LGL network}
T-cell large granular lymphocytic (T-LGL) leukemia is a disease that is caused by activated cytotoxic T-cells surviving and possibly proliferating instead of undergoing activation induced cell death. In \cite{Zhang}, Zhang et al. constructed a signal transduction network of activation induced cell death and its deregulation in T-LGL leukemia, and modeled it with a Boolean model of this process. Here, we use the version of the model in \cite{Zanudo3}. This T-LGL network contains 60 nodes and 141 edges; we identified several near-minimal FVSs, all of size twelve. The model has three attractors, two survival attractors with overlapping node states that include the OFF state of the node Apoptosis, and one apoptosis attractor which includes the ON state of the node Apoptosis. Using our method, we identified multiple FVS subsets; in this follow-up analysis we aim to evaluate which of the $2^L$ interventions for each FVS subset drives the system out of the survival attractors and into the apoptosis attractor. As a state with Apoptosis=1 reflects the biological commitment to apoptosis even if it is not identical to the system's apoptosis attractor, we can analyze a FVS subset's effectiveness either by the state of Apoptosis after each simulation or by the \textit{To Control} and \textit{Away Control} values.

The top five single node FVS subsets predicted by the propagation intersection metric were IL2RB, NFKB, RAS, S1P, and TBET. The intervention expected to be successful in driving the system into the apoptosis attractor is to set the target node into its state corresponding to the apoptosis attractor. This means the OFF state for S1P. Indeed, knockout of S1P drives the system away from the survival state and into the apoptosis attractor a majority of the time (Table \ref{tab5}). However, the situation is less clear for the other four nodes in this list because they are ON in all attractors; these nodes are partially informative when used alone and become fully informative when used in combination with S1P. After simulating both the knockout and constitutive expression of each of these four nodes in the Boolean model, we found that knockouts of NFKB, IL2RB, or RAS increases the basin of the survival state to \textgreater 90\%. The resulting attractor is not close to the apoptosis attractor but it does have the Apoptosis node in the ON state (Table \ref{tab5}). The effectiveness of S1P, NFKB, and RAS knockout in a practical setting can be corroborated with experimental evidence \cite{Layek,Shah,Epling}. TBET has a smaller Apoptosis basin of attraction than the other four identified interventions, however it did drive the system to new attractors that are reasonably different from the original survival attractor and have Apoptosis oscillating. We also simulated the fully informative single node intervention with the next highest percentile cutoff, FLIP, and found that knocking it out does not meaningfully change the basins compared to the unperturbed (wild-type) system. This indicates that fully informative FVS subsets might not always be better options than partially informative ones in terms of their ability to control the network.

\begin{table*}
\caption{\label{tab5}Top five FVS subsets of size 1, 2, and 3 for the T-LGL network using the propagation intersection metric. For each FVS subset, we identified the interventions that drive the system to the apoptosis state. The success of the interventions is measured by their \textit{To Control} value, \textit{Away Control} value, and by the state of the Apoptosis node averaged over 100 simulations. All one node interventions increase the basin of the apoptosis attractor when compared to the unperturbed model, but only S1P fully drives the system to the wild-type apoptosis attractor. The other one node interventions drive the system away from the wild-type attractors to new survival and apoptosis attractors. Combining the intervention state for two one-node interventions that were successful is not always a successful two-node intervention. For example, separately fixing NFKB OFF or S1P OFF fully drives the system into the apoptosis state, but the combined intervention \{NFKB = 1, S1P = 0\} performs better than combining the best individual interventions as \{NFKB = 0, S1P = 0\} (indicated using an asterisk). After determining the intervention states that were the most successful in activating Apoptosis for each subset, we found that the two- and three-node FVS subsets were able to meaningfully increase the Apoptosis basin of attraction compared to the included single interventions.}
\begin{ruledtabular}
\begin{tabular}{l c r r r r r}
Intervention Set&\shortstack{Best\\intervention\\to induce\\apoptosis}&\shortstack{\textit{To}\\\textit{Control}}&\shortstack{\textit{Away}\\\textit{Control}}&\shortstack{Size of Basin\\w/ Apoptosis\\OFF}&\shortstack{Size of Basin\\w/ Apoptosis\\ON}&\shortstack{Size of Basin\\w/ Apoptosis\\oscillating}\\
\hline
Wild-Type&N/A&N/A&N/A&44\%&56\%&0\%\\
S1P&0&0.89&1.00&0\%&100\%&0\%\\
NFKB&0&0.00&1.00&10\%&90\%&0\%\\
IL2RB&0&0.00&0.96&0\%&100\%&0\%\\
RAS&0&0.00&1.00&7\%&93\%&0\%\\
TBET&0&0.00&1.00&37\%&0\%&63\%\\
FLIP (next fully informative node)&0&0.08&0.08&44\%&56\%&0\%\\
NFKB, S1P&10&1.00&1.00&0\%&100\%&0\%\\
NFKB, S1P*&00&0.00&1.00&16\%&84\%&0\%\\
NFKB, RAS&10&0.00&1.00&0\%&100\%&0\%\\
NFKB, GRB2&00&0.00&1.00&13\%&87\%&0\%\\
NFKB, IL2RB&01&0.00&1.00&0\%&100\%&0\%\\
JAK, S1P&10&0.93&1.00&0\%&100\%&0\%\\
NFKB, S1P + one of&101&1.00&1.00&0\%&100\%&0\%\\
\{IL2RB, RAS, BID, TBET, GRB2\}\\
\end{tabular}
\end{ruledtabular}
\end{table*}

We also identified the top 5 two and three node FVS subsets of the propagation intersection metric. We simulated all four (for two nodes) or eight (for three nodes) combinations of knockouts and constitutive expressions for each subset. We identified the intervention that best drove the system to the apoptosis state and recorded this intervention in Table \ref{tab5}. The two node interventions drive the system to an apoptosis state more consistently than the single node interventions, but many of the interventions were not driving the system to the wild-type apoptosis attractor. 

\begin{figure}
\includegraphics[width=\linewidth]{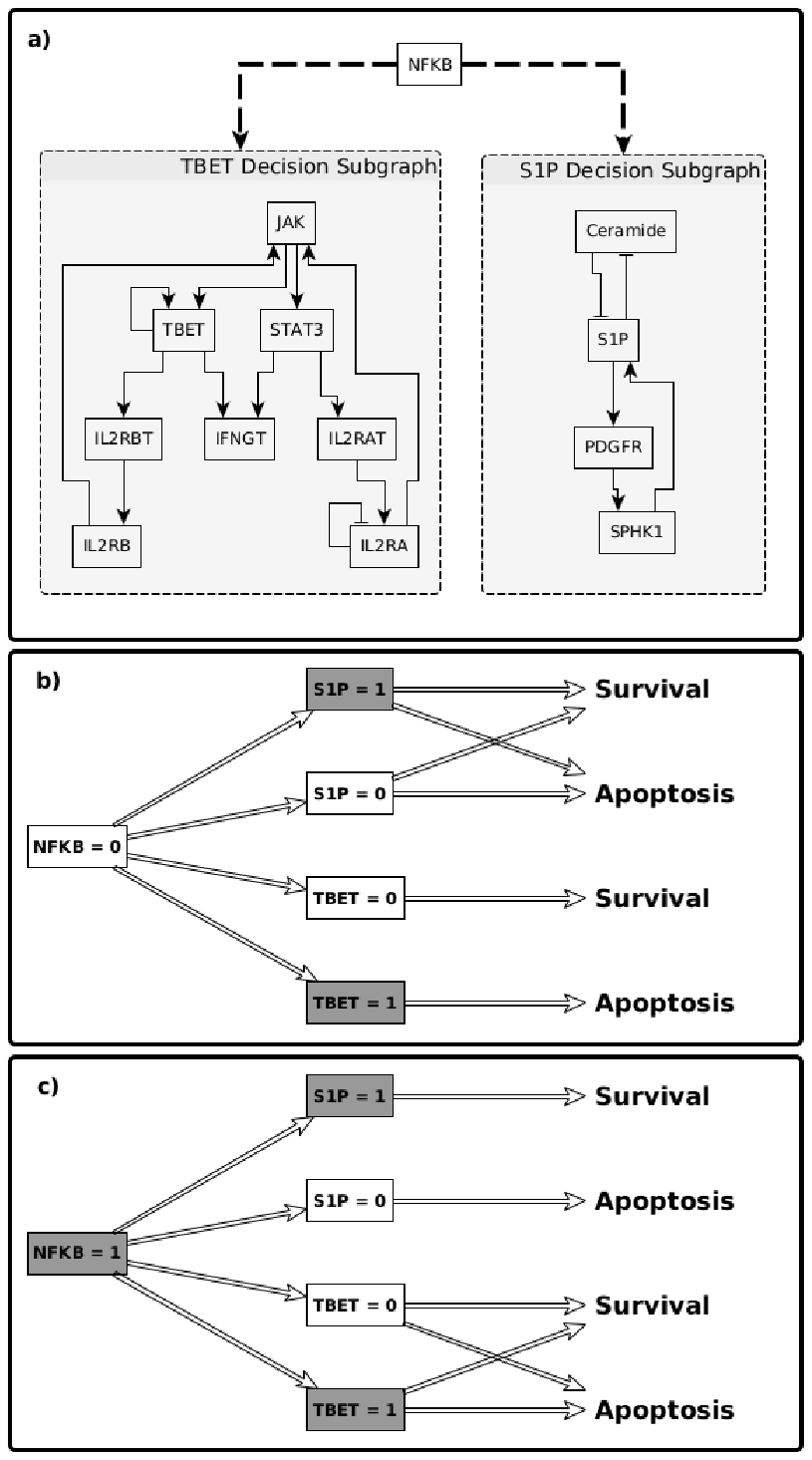}
\caption{\label{fig10}Illustration of how NFKB, S1P and TBET affect the attractor of the T-LGL network. (a) The subgraphs of the network that determine which state the system will be driven to. Depending on the state of NFKB, the S1P subgraph or the TBET subgraph plays a key role in determining the attractor. (b) When NFKB is OFF, only the TBET subgraph determines if the network will enter the Survival or Apoptosis states. Fixing the S1P subgraph to either state does not determine if the system will enter either attractor. (c) Conversely, when NFKB is ON, which is its wild-type state in all the attractors, the S1P subgraph determines which state the network enters, and the TBET subgraph cannot determine the final state.}
\end{figure}

Furthermore, the best performing combined intervention is not always a combination of the best performing single interventions. For example, the intervention set \{NFKB = 1, S1P = 0\} is more effective than the set \{NFKB = 0, S1P = 0\}. Once NFKB = 1 and S1P = 0 are locked in, the system will almost always be driven to the wild-type apoptosis attractor. Upon further analysis of the Boolean model, we discovered the mechanisms that determine which state of NFKB will be more effective. Depending on the state of NFKB, two different subgraphs determine the attractor [Fig. \ref{fig10}(a)]. Certain state configurations of these subgraphs are stable motifs, meaning that these state configurations can be maintained regardless of the rest of the network. Stable motifs' role in driving a Boolean system to its attractors has been studied in detail \cite{Zanudo2,Zanudo3}. In the wild-type system, NFKB naturally turns ON, so the S1P subgraph determines the attractor [Fig. \ref{fig10}(b)]. Turning S1P OFF drives the system into the Apoptosis state. Conversely, when NFKB is fixed OFF, the TBET subgraph determines the state of the system [Fig. \ref{fig10}(c)]. Fixing TBET, IL2RB, or JAK ON drives the system to the Apoptosis state. However, fixing S1P OFF cannot return the system to the original wild-type apoptosis attractor and instead drives to new survival and apoptosis attractors. This example illustrates why the state of the nodes in a FVS subset with the highest ability to drive the system towards or away from an attractor of interest does not always correspond to the one matching the nodes states of the target attractor. In cases with nodes whose state is the same in every attractor, such as NFKB, either fixing the node in the state of the attractors or in the state opposite the attractors could be better at achieving high control values depending on the model.

In the three node case, setting every node in the FVS subset to its state in the wild-type apoptosis attractor is the most successful intervention for reaching an attractor with Apoptosis ON. The top ranking FVS subsets almost all achieve a \textit{To Control} value of 1 and all all achieve an \textit{Away Control} value of 1, and every intervention contains the combination \{NFKB = 1, S1P = 0\}. Thus, when any node in the subset is distinct between attractors, such as S1P, the ambiguity in choosing the states for other nodes in the subset that are not distinct between attractors, such as NFKB, is no longer an issue.

\subsubsection{\label{subsubsec:FABRCA}Fanconi anemia/Breast Cancer (FA/BRCA) network}
This Boolean network models the repair process of a cell following different types of DNA damage: interstrand cross-links (ICLs), double-strand breaks (DSBs), and DNA adducts (ADDs). In the unperturbed model from Rodr\'{i}guez et al. in \cite{Rodriguez} there is only one attractor: a cell cycle attractor in which the node CHKREC oscillates and the rest of the nodes are off. However, they also analyzed the attractors of the model in the presence of different mutations, with specific nodes either constitutively expressed or knocked-out as a consequence of the mutation studied. Here, we apply our method to a variant of the model where BRCA1 is knocked out (variant \#3). The mutant has the original cell cycle attractor and also has a DNA damage attractor marked by the activation of the node DSB. The minimal FVSs we identified for this network have eight nodes, which include DSB and the node that indicates the cell cycle (CHKREC). We do not consider these two nodes as possible intervention targets because they directly determine the resulting attractor. In addition to the original cell cycle and DNA damage attractors, some interventions create two new attractors types: attractors without DNA damage that also don't undergo the cell cycle, and attractors where the cell cycle persists despite sustaining DNA damage. Along with determining the \textit{To Control} and \textit{Away Control} values relative to the two original attractors, we also determined the basin of attraction of attractors that show no DNA damage. We use our percentile cutoff method to identify the three top-performing one, two, and three node FVS subsets in the all metric intersection metric(see Table \ref{tab6}). All of these FVS subsets are fully informative, so determining the intervention states for each node in the FVS subset is more straightforward than in the previous example.

\begin{table*}
\caption{\label{tab6}Top three FVS subsets of size 1, 2, and 3 for the BRCA1 knockout mutant of the FA/BRCA network using the propagation intersection metric. ATM+ indicates all interventions that include ATM will always drive the system to the correct state if all nodes are in the state of that attractor. Furthermore, we also include the single-node knockout of ssDNARPA for comparison to the larger identified subsets that did include ssDNARPA.}
\begin{ruledtabular}
\begin{tabular}{l c r r r r}
Intervnetion Set&\shortstack{Best\\intervention\\to enter\\Cell Cycle}&\shortstack{\textit{To}\\\textit{Control}}&\shortstack{\textit{Away}\\\textit{Control}}&\shortstack{Size of Basin\\w/ DSB\\OFF}&\shortstack{Size of Basin\\w/ DSB\\ON}\\
\hline
Wild-Type&N/A&N/A&N/A&40\%&60\%\\
ATM&0&1.00&1.00&100\%&0\%\\
ATR&0&0.25&0.34&55\%&45\%\\
ssDNARPA (for completeness)&0&-0.11&0.14&41\%&59\%\\
ICL&0&0.00&0.00&51\%&49\%\\
ATM+&0+&1.00&1.00&100\%&0\%\\
ATR, ICL&00&0.554&0.554&59\%&41\%\\
ICL, ssDNARPA&00&0.26&0.41&53\%&47\%\\
&01&-0.17&0.36&60\%&40\%\\
ATR, ICL, ssDNARPA&000&0.44&0.56&57\%&43\%\\
&001&0.00&0.00&75\%&25\%
\end{tabular}
\end{ruledtabular}
\end{table*}

Knocking out the FVS node ATM can drive the system out of the DNA damage attractor, so any FVS subsets that include ATM can also drive the system out of the DNA damage attractor. Knockouts of ATR or ICL do not perform as well as ATM, but they do decrease the basin size of the DSB ON attractors compared to the system without an intervention. The knockout of ICL targets both attractors, so it is uninformative, resulting in \textit{To Control} and \textit{Away Control} values of zero, but it does slightly improve the basin of the DSB OFF attractor. For completeness, we also simulated the knockout of the node ssDNARPA as it appears in some of the identified two- and three-node subsets, but individual knockout of ssDNARPA does not significantly influence the basin of the DSB ON attractors. Interventions of larger subsets that include ATR, ICL, or ssDNARPA do show improvement with respect to the single node cases, resulting in a larger decrease in the basin of the DNA damage attractor. Interestingly, while ssDNARPA is OFF in the cell cycle attractor, in some larger subsets, fixing ssDNARPA ON deactivates DSB more often than fixing ssDNARPA OFF, but either state of ssDNARPA in a multi-node intervention decreases the basin of the DNA damage attractor more than the single node intervention of ssDNARPA. For two and three node FVS subsets, we note that without controlling ATM, the basin of the DSB OFF attractor cannot exceed a maximum size around 80\%, highlighting the importance of ATM in controlling this variant of the FA/BRCA network.

\section{\label{sec:Disc}Discussion}
We identified seven topological metrics that had the potential to sort FVS subsets by their ability to drive the dynamic trajectory of a biological network. Verifying the effectiveness of each intervention in multiple Boolean models of biological networks, we determined the predictive power of each topological metric. We found that the centrality measures performed adequately. This moderate performance may be because centrality measures only glean local or path-based information from the network and ignore the cycles, which are crucial for understanding the multistability in the network. To our surprise, we found that the cycle-based metrics performed poorly. This indicates that only using the cycle structure of the network does not successfully differentiate between FVS subsets. This poor predictive power may be because the FVS already implements knowledge of the network's cycle structure, and reusing the cycle structure in the cycle-based metrics results in diminishing returns. We demonstrated that the three propagation metrics, which utilize both path information and cycle information, are the best at ranking FVS subsets based on their ability to control the network.

We further demonstrated that taking an intersection of the top ranking subsets of these three propagation metrics is even more successful. Using these intersections, we focused on the high precision and low recall regime because we expect FVS subsets with high intersection metric values to be sufficient but not necessary for having a high control value. While an intersection of all seven metrics captures more structural properties of the network than the propagation metrics, the all metric intersection metric is often too strict because it includes topological metrics with little predictive ability. Based on the analysis of the studied Boolean models, we find that the propagation intersection metric is less stringent, more precise, and identifies more intervention FVS subsets than the all metric intersection metric in the high precision and low recall regime.

To explore how well the propagation intersection metric identifies FVS subsets with high control value, we simulate the top ranking FVS subsets identified by the propagation intersection metric. We found that the top ranking FVS subsets have high control values for two arrays of Boolean networks: the array of networks we used to study our topological and intersection metrics and a second array of networks. We also verified that these identified FVS subsets perform better than randomly chosen node subsets on the same two arrays of networks. Our in-depth analysis of the interventions predicted for two specific models led to general insights. Looking at the T-LGL network, we highlight an ambiguity when trying to identify the intervention state with the highest control value for nodes that have the same state in every attractor. We find that this ambiguity is eliminated when a node whose state is distinct between attractors is added to the intervention. Looking at the 3rd variant of the FA/BRCA network, we confirm that if a FVS subset performs well, then supersets of this subset tend to achieve control values that are equal or greater than the original FVS subset. Using this analysis, we support our hypothesis that the top ranking FVS subsets identified by the propagation intersection metric are successful at driving networks to and away from their attractors. 

In summary, our approach successfully identifies few-node combinatorial interventions that can drive a network into a desired attractor and away from undesired attractors. For many of the networks in our study we were able to identify two or three-node interventions that controlled the network. The largest benefit of this approach is that it is based solely on the topology of the system, so it can be used for systems whose dynamics are poorly characterized or to identify interventions that are robust to the dynamic details of the system. Our method has a wide breadth of applications among biological as well as non-biological networks. Being able to identify crucial, attractor-driving nodes in dynamical systems has practical implications such as identifying targets of restorative or preventative interventions. Overall, our study contributes to the body of work documenting how certain dynamical behaviors of a network model, and the methods to elicit them, can be fully determined given only topological knowledge of the network.

\section{\label{sec:Methods}Methods}
First, we introduce key notations. $N$ denotes the number of nodes in the network, and $M$ denotes the number of attractors. Individual attractors are denoted $A_i$, where $i$ goes from $1$ to $M$. $n$ denotes an individual node in the network. The state of node $n$ in attractor $i$ is represented as $s_{ni}$. $S\subseteq FVS = \{n_1, n_2, \dots n_L\}$ denotes a FVS subset of size $L$. $P = \{p_1(n_1), \dots, p_L(n_L)\}$ denotes an intervention of a FVS subset of size $L$. Here $p_j(n_j)$ denotes an individual intervention on node $n_j$ (e.g, $P = \{p_1(S1P), p_2(NFKB)\} = \{0, 1\}$ means S1P is being driven to its OFF state and NFKB is driven to it's ON state).

\subsection{\label{subsec:Bool}Boolean modeling}
A Boolean model is a discrete dynamic model that characterizes each node of a network with a state variable that can take one of two values. These values are 0 and 1, interpreted as OFF (inactive) and ON (active), respectively. Each node is characterized with an update function usually expressed by logical operations. This update function takes as input the state values of other nodes in the network, and its output determines the node's state at the next time step. We use a stochastic asynchronous update scheme in which at every time step a randomly selected node's state is updated by evaluating its update function. An asynchronous update scheme updates nodes individually and in a stochastic manner, as opposed to a synchronous update.

\subsection{\label{subsec:TopoMets}Identification of FVS subsets using topological metrics}

Given that minimal and near-minimal FVS are non-unique \cite{Zanudo, Galinier, Mochizuki}, we identify multiple near-minimal FVS, obtaining a superset of every node that could be in any of the near-minimal FVS. When finding multi-node subsets, we only analyze subsets that are part of the same near-minimal FVS. In other words, we ignore subsets that contain nodes that appear in separate near-minimal FVSs.

In the following we describe each topological metric.

\subsubsection{\label{subsubsec:OutDegree}Sum of out-degrees}
The sum of the out-degrees is a local measure that measures the number of nodes the subset is directly connected to.
\begin{equation}
\label{equ:OutDegree}
Out Degree(S) = \sum_{i = 1}^{L} Out Degree(n_i)
\end{equation}

\subsubsection{\label{subsubsec:Distance}Distance}
The distance from node $n_i$ to node $n_j$ ($l_{ij}$) is a measure of how close these nodes are to each other, but to account for unreachable nodes the inverse distance ($\frac{1}{1+l_{ij}}$) is used. The average of these inverse distances from a node to every other node in the network defines the distance metric. When working with a subset of more than one node ($L > 1$), the distance ($l_{ij}$) used in the equation is the distance to the closest node in the subset, so it is defined as:
\begin{equation}
\label{equ:Distance}
Average Inverse Distance (S) = \frac{1}{N}\sum_{i = 1}^{N} \min_{n_j \in S}\frac{1}{1+l_{ij}}
\end{equation}
Edge sign is not incorporated into the distance metric, thus it cannot distinguish between positive and negative paths.

\subsubsection{\label{subsubsec:PRINCE}PRINCE propagation}
The PRINCE Propagation metric reflects how much information from a node propagates through the system. It is formulated by applying a constant perturbation on the specified node subset and then allowing the perturbation's information to be distributed to the node's neighbors according to a discrete time dynamic process. This initial perturbation vector ($\overrightarrow{Y}$) is a discrete value vector of size ($N$), where the value corresponding to each node is either 1 (fixing the node ON), 0 (no effect on the node), or -1 (fixing the node OFF).

The value propagated through an edge depends on the edge's sign, i.e., a negative edge propagates a negative value. This dynamic process attributes every node a value between -1 and 1. The value is the node's information score, which represents how much information from the perturbation is received. An absolute value of 1 means we have full information about the node, while a value of 0 means that we do not know anything about the given node.

The information scores of every node in the system are based on a normalization scheme and a propagation dropoff value ($\alpha$). The normalization scheme mimics mass flow. The out-going edges split the information evenly between every successor. The amount of intake is also distributed evenly to every predecessor. The propagation dropoff value causes the information from an input node to decrease the farther it travels. The propagation is calculated based on the algorithm:
\begin{equation}
\label{equ:PRINCE}
\pi(t+1) = \alpha\textbf{W}^{\prime T}\pi(t)+(1-\alpha)\overrightarrow{Y}
\end{equation}
where $\pi(t)$ is the propagation vector at time t, $\textbf{W}^\prime = D_1^{-1/2}\textbf{W}D_2^{-1/2}$ is the normalized adjacency matrix, in which $D_1$ is a diagonal matrix where element $D_1(i, i)$ is the out-degree of node $i$, $D_2$ is a diagonal matrix where element $D_2(i,i)$ is the in-degree of node $i$, and $\textbf{W}$ is the signed adjacency matrix where element $W_{ij}$ is 1 if there is a positive edge from node $i$ to node $j$, -1 if the edge is negative, and 0 if there is no edge between nodes $i$ and $j$. $\alpha$ is the propagation dropoff value, which we set to $\alpha=0.9$ for our calculations following the value used in \cite{Santolini}. 

The long term behavior of the system can be characterized by its steady state $\pi^*$, which is such that $\pi^* = \alpha\textbf{W}^{\prime T}\pi^*+(1-\alpha)\overrightarrow{Y}$. To calculate $\pi^*$, we follow \cite{Santolini} which shows $\pi^* = (1-\alpha) \overrightarrow{Y} (\textbf{I}-\alpha\textbf{W}^{'T})^{-1}$. To measure how much information we have on every node, the absolute value of $\pi^*$ is taken. The PRINCE Propagation value for a specific subset is the average of these absolute values of the information scores taken over every node in the network ($\overline{\pi^*})$.

Because the absolute values of the information scores are taken, an intervention of 1 and -1 return the same value in a one-node intervention, so only the inputs of 1 are tested. When considering a multi-node perturbation, some further modifications need to be used. Instead of a constant input of 1 to every node in the subset, all combinations of inputs of 1 and -1 are taken, and the maximum of these results is used as the PRINCE value for this subset.

\subsubsection{\label{subsubsec:ModPRINCE}Modified PRINCE propagation}
We propose a variant of the PRINCE Propagation algorithm that only normalizes the propagation by the in-degree of each node. This better treats the initial perturbation as information instead of mass, so it can fully spread to all successors instead of being split among them. To keep the system convergent, the adjacency matrix is changed to $\textbf{W}^\prime = \textbf{W}D_2^{-1}$. Other than adjusting the adjacency matrix, the values for this metric are obtained the same way as in the original PRINCE propagation. It still accounts for edge signs by allowing for the propagation value to be negative, and it also still treats multi-node interventions by considering every combination of input values.

\subsubsection{\label{subsubsec:CheiRank}CheiRank}
CheiRank is an extension of PageRank, which consists of a modified random walk over the nodes of a network. CheiRank follows the same algorithm as PageRank but traverses the network's edges in reverse. The random walk is augmented with a probability $\alpha$, where there is a probability of $1-\alpha$ that the walker jumps to a random node instead of following an edge. The probability vector $\pi(t)$ of the random walk follows:
\begin{equation}
\label{equ:CheiRank}
\pi(t + 1) = \alpha\textbf{A}^T\pi(t) + (1-\alpha)\frac{\textbf{E}}{N_n}\pi(t)
\end{equation}
where $\textbf{A}$ is the adjacency matrix of the network with each row normalized so it sums to 1, so there is an equal probability to travel to any neighbor nodes, and $\textbf{E}$ is an all 1 matrix.

In the long-time limit, the system will reach a unique steady state $\pi^*$, that satisfies $\pi^* = [\alpha\textbf{A}^T\ + (1-\alpha)\frac{\textbf{E}}{N_n}] \pi^*$, where each value in $\pi^*$ is the CheiRank of that node and it is the probability that the random walker is found in this specific node. Because CheiRank denotes a probability, when the subset has multiple nodes in it, the probabilities are combined multiplicatively in the following way: $CheiRank(n_0, \dots n_L) = 1-(1-CheiRank(n_0))*\dots*(1-CheiRank(n_L))$

\subsubsection{\label{subsubsec:Cycle-based}Cycle-based metrics}
The cycle-based metrics estimate the multistability reintroduced into the system after the FVS is removed and a subset of it is reintroduced. We used two topological measures as an estimate of multi-stability: the number of positive cycles (cycles with an even number of negative edges), and the size of the strongly connected component. Both measures return 0 for subsets containing nodes that are not in the FVS.

\paragraph{\label{ppg:Cycles}Number of Positive Cycles}
The number of positive cycles of a FVS subset ($S$) is defined as the number of positive cycles left in the network after the FVS has been removed and the subset has been reintroduced:
\begin{equation}
\label{equ:Cycles}
\#Cycles(S) = \# Positive Cycles(G \backslash FVS+S)
\end{equation}

\paragraph{\label{ppg:SCC} Size of the Strongly Connected Component}
The SCC of a FVS subset ($S$) is defined as the size of the largest SCC after the FVS has been removed and the subset has been reintroduced:
\begin{equation}
\label{equ:SCC}
SCC Size(S) = Size\;of\;Largest\;SCC(G \backslash FVS+S)
\end{equation}
Unlike the cycles, the signs of the edges aren't incorporated in this metric.

\subsection{\label{subsec:BasinSize}Calculating the basin of attraction of the wild-type attractors with or without interventions}
To determine the attractors of the wild-type (unperturbed) system, we use the trapspace method in the BioLQM toolkit \cite{Naldi2}. Technically, the objects identified by this method are the minimal trap spaces (subsets of the state space specified by fixing the value of a set of nodes which if entered cannot be exited), not the attractors, but there is often a one-to-one correspondence between them, and trap spaces can be identified efficiently \cite{Pauleve}.The attractors include fixed points (steady states) in which the state of all the nodes is fixed and complex attractors, in which the system keeps revisiting a finite set of states. We calculate each attractor's basin of attraction by performing 1000 simulations starting from random initial conditions and using a general asynchronous update. The simulation continues until an attractor is reached, and we count how many of the simulations reach each of the system's attractors to determine the basins of attraction.

For each intervention, we perform 100 simulations from random initial conditions using a general asynchronous update. After each simulated intervention, we measure how close the final state of the system, $Sim(\overrightarrow{P})$, is to the attractors of the wild-type system, $A_i$. We measure the closeness to each attractor through a normalized Hamming distance. The Hamming distance $h(S_1, S_2)$ measures the fraction of nodes with different values between two different states $S_1$ and $S_2$. The normalized Hamming distance between the final state and every wild-type attractor is used to generate an attractor control value ($C$) for each attractor defined as:

\begin{align}
\label{equ:Control}
&C(h(A_i, Sim(P));F) = \notag \\
&\left\{\begin{array}{ll}1 - \frac{h(A_i, Sim(P))}{F} & \mbox{if } h(A_i, Sim(P)) \leq F \\0 & \mbox{if } h(A_i, Sim(P) > F\end{array}\right.
\end{align}

The attractor control value is a linearly decreasing function with a cutoff $F$, which indicates when the state is too far away from our attractor. The $F$ parameter is defined to be halfway between the two closest attractors in the network, $F = \frac{\min_{i,j}h(A_i, A_j)}{2}$. This ensures that each final state in a simulation is attributed to one attractor and the other attractors get a value of 0 for that specific simulation.

For each simulation, we get an attractor control value for every attractor, but only one is nonzero. These values are averaged over the 100 simulations with different, random initial conditions. The average for each attractor is considered the basin of attraction for that attractor in the perturbed system, $B(A_i|P) = \overline{C(h(A_i, Sim(P));F)}$. The intervention basin vector is defined as $\overrightarrow{B(P)} = (B(A_1|P), B(A_2|P), \dots, B(A_M|P))$. As a point of comparison, we consolidated the wild-type basin values into the wild-type basin vector, $\overrightarrow{WTB} = (WTB(A_1), WTB(A_2), \dots, WTB(A_M))$, where $WTB(A_i) = B(A_i|0) = \overline{C(h(A_i, Sim(0)); F)}$ is the wild-type basin of each attractor, which is averaged over 1000 simulations.

\subsection{\label{subsecc:ToAwayControl}Calculation of \textit{To Control} and \textit{Away Control}}
To study the effects of an individual intervention, we separated the wild-type system's attractors into two categories: Target attractors and Non-Target attractors. These two categories are uniquely determined based on each specific intervention. A Target attractor is an attractor wherein the intervention state of each node is the same as the node's state in the attractor, and a Non-Target attractor is an attractor wherein the intervention state of any node is not in the attractor's state.

We describe this using indicator functions. The indicator function $I(s_{{n_j}i}, p_j(n_j))$ compares the state of node $n_j$ in a specific attractor, $s_{n_ji}$, to the state the node is fixed into through the intervention, $p_j(n_j)$. If they are the same, the function returns 1 and if they are different the function returns 0, so it is defined as:

\begin{equation}
\label{equ:IndicatorFunc}
\textbf{I}(s_{n_ji}, p_j(n_j)) = \left\{\begin{array}{ll} 1 & \mbox{if } s_{n_ji} = p_j(n_j) \\0 & \mbox{if } s_{n_ji} \neq p_j(n_j) \end{array}\right.
\end{equation}

To extend this to multiple nodes, the indicator functions for every node in the intervention are multiplied together. These products are merged into a single identification vector of size $M$, $\overrightarrow{\textbf{I(P)}} = \{\prod_{j = 1}^L[\textbf{I}(s_{n_j1}, p_j(n_j))],\dots,\prod_{j = 1}^L[\textbf{I}(s_{n_jM}, p_j(n_j))]\}$. The attractors for which the corresponding elements of the identification vector are 1 are the Target attractors and the attractors for which the elements are 0 are the Non-Target attractors.

After labeling each attractor as a Target or Non-Target, we filter out interventions that are considered uninformative. Interventions where every attractor is a Target attractor are uninformative because we cannot attribute changes in an attractor's basin to the intervention. Similarly, the \textit{To Control} metric for an intervention is uninformative if every attractor is considered a Non-Target. Because we are not driving to any attractor, we cannot measure how often we reach a Target attractor, but we can still measure if the intervention drives the system away from the attractors of the system, so we denote these interventions as partially informative. We denote an intervention as fully informative if the intervention has attractors that are Targets and attractors that are Non-Targets.

For every (fully or partially) informative intervention, we sum the wild-type basins of every Target attractor into one wild-type Target basin, $WTTB(P) = \overrightarrow{I(P)} \bullet \overrightarrow{WTB}$, and sum the basins for every Non-target attractor into one wild-type Non-Target basin, $WTNTB(P) = 1-WTTB(P) = (\overrightarrow{1} - \overrightarrow{I(P)})\bullet\overrightarrow{WTB}$. Similarly, we combine the intervention attractor's basins into an intervention Target basin, $TB(P) = \overrightarrow{I(P)}\bullet\overrightarrow{B(P)}$, and an intervention Non-Target basin, $NTB(P) = 1-TB(P) = (\overrightarrow{1} - \overrightarrow{I(P)})\bullet\overrightarrow{B(P)}$.

We then find the percent difference between these basins to understand how the intervention affects the system. The difference between the intervention Target basin and wild-type Target basin is the intervention-specific \textit{To Control}, and the difference between the intervention Non-Target basin and wild-type Non-Target basin is the intervention-specific \textit{Away} Control. Correspondingly, the \textit{To Control} and \textit{Away Control} values range between -1 and 1.

\begin{align}
\label{equ:ToControlPert}
&TC(P) =\notag\\
&\left\{\begin{array}{ll}\frac{TB(P) - WTTB(P)}{1-WTTB(P)} & \mbox{if } TB(P) \geq WTTB(P)) \\\frac{TB(P) - WTTB(P)}{WTTB(P)} & \mbox{if } TB(P) < WTTB(P)\end{array}\right.
\end{align}
\begin{align}
\label{equ:AwayControlPert}
&AC(P) =\notag\\
&\left\{\begin{array}{ll}\frac{WTNTB(P) - NTB(P)}{1-WTNTB(P)} & \mbox{if } NTB(P) \geq WTNTB(P)) \\\frac{WTNTB(P) - NTB(P)}{WTNTB(P)} & \mbox{if } NTB(P) < WTNTB(P)\end{array}\right.
\end{align}

Once all $2^L$ possible interventions are tested or filtered out for a subset, we aggregate the values into a singular aggregate \textit{To Control} and \textit{Away Control} values for the entire subset by taking the maximum of all available values. Taking a maximum of all the $2^L$ interventions ensures that the aggregate \textit{To Control} and \textit{Away Control} values of the entire FVS is 1, indicating full control over the network.

\begin{equation}
\label{equ:ToControl}
TC = \max_PTC(P)
\end{equation}

\begin{equation}
\label{equ:AwayControl}
AC = \max_PAC(P)
\end{equation}

\subsection{\label{subsec:LogReg}Logistic regression and AUPRC}
Logistic regressions are a type of binary classifiers. To convert our continuous FVS subset control values to a binarized value of a successful versus unsuccessful FVS subset we impose a threshold. We used a threshold of 0.9 to binarize the control values and found there were no major changes in our results using thresholds in the range 0.5-0.95. The logistic regression fits a logistic function to the binarized data and results in a curve in which the x-values of the fit are the values of the topological metrics and the y-values indicate the probability of that topological value resulting in successful control (i.e., control above 0.9). Therefore, the y-value for the logistic function is between 0 and 1, and in an ideal situation, it starts at 0 for low x-values, increases to 1 as the x-value increases, reflecting how we expect topological values to be positively correlated with control.

This regression can be modified by applying weights to the negative and positive data points. By default, we create an unbalanced regression which weights every data point equally. We also test a balanced regression where data points are weighted by the number of points in their data set, i.e., positive data points are weighted by the relative number of positive data points and negative data points are weighted by the relative number of negative data points. This means the positive data set and negative data set are weighted to correct for differences in their relative sizes.

We use the precision-recall curve and the area under the precision-recall curve (AUPRC) to evaluate the strength of the fit. We choose the AUPRC over the more common receiver operating characteristic (ROC) curve because there is often a large imbalance in the binarized control values, which limits the usefulness of the ROC curve for our analysis \cite{Davis}. To generate a precision-recall curve, the entire range of a topological value is scanned. As each value is scanned over, the value of the logistic regression function is used to sort the data into four categories: true positives (TP), false positives (FP), false negatives (FN), and true negatives (TN). Then based on these values, a precision ($\frac{TP}{TP+FP}$) and recall ($\frac{TP}{TP+FN}$ value can be calculated. As the precision and recall are calculated, they are plotted on a curve, so when the entire x-axis is scanned, the precision-recall curve is formed. The AUPRC is calculated from the precision-recall curve and consolidates the curve into a single value that can indicate how well the topological metric sorts the data.

The AUPRC is a positive value less than or equal to 1, where 1 indicates that the topological metric perfectly sorts between positive and negative binarized control values. When the AUPRC value is equal to the fraction of positive data points, it indicates that the topological metric does not sort the FVS subsets better than random. As performing better than random is not sufficient to confirm that the topological metric can predict which FVS subsets are successful, we introduced an AURPC predictive threshold which is halfway between the random expectation and the maximum score of 1 ($1-\frac{1-PositiveFraction}{2}$)

\subsection{\label{subsec:Intersections}Intersections}
For each topological metric, the FVS subsets are ordered in increasing order of the value of the metric. Then, this ordered list is converted into percentile values $f(S,m)$ for each FVS subset $S$. For a set of topological metrics of interest, the sets of FVS subsets above a fixed threshold percentile $pc$ are intersected. We refer to metrics that intersect the top ranking FVS subsets of a set of topological metrics as an intersection metric, where each intersection metric is denoted by their set of topological metrics.

\begin{equation}
\label{equ:Intersection}
\begin{split}
IM(pc&|Topological Metrics) =\\
&\bigcup_{m \in Topological Metrics} f(S, m) \geq pc
\end{split}
\end{equation}

We investigated three different intersection metrics: an intersection of all seven metrics, an intersection of only the propagation metrics, and an intersection of the CheiRank and Modified PRINCE metrics.

In an intersection metric, we assign to each FVS subset a percentile cutoff value $IP(S,Metrics)$ that is equal to the minimal percentile cutoff value of the FVS subset among the metrics being intersected. This is equivalent to choosing the maximal $pc$ value for which $S$ is found in $IM(pc|TopologicalMetrics)$ given the chosen metrics.
\begin{align}
\label{equ:IntPercentile}
\begin{split}
IP(S&,Topological Metrics)\\
&= \min_{m \in Topological Metrics} f(S,m)\\
&= \max_{pc}[S \in IM(pc|Topological Metrics)]
\end{split}
\end{align}

\subsection{\label{subsec:Computational}Computational implementation}
The methods were implemented in python using various libraries and modules. The code is available at \url{https://github.com/EliNewby/FVSSubsets}. This code consists of two python files: \texttt{FindBestSubsets.py} and \texttt{RunSimulations.py}. \texttt{FindBestSubsets.py} finds the topological metric values and intersection metric values for the FVS subsets of a network topology and uses these values to rank the FVS subsets. \texttt{RunSimulations} takes a list of node sets and a Boolean model for a network and calculates the \textit{To Control} and \textit{Away Control} values for each node set. We also include a Jupyter notebook that includes an example showing the output of these two functions on the T-LGL network and that also reproduces the plots included in the figures of this manuscript.

To identify the near-minimal FVS and its subsets, a python code developed by Gang Yang \cite{Zanudo} was used, which utilizes the simulated annealing algorithm presented in \cite{Galinier}. This code is available at \url{https://github.com/jgtz/FVS_python3}. NetworkX was used to analyze the structure of the networks and to calculate the values of the topological metrics. NetworkX implemented some of the topological metrics we used. Custom Python code and NetworkX functions were used for the ones that were not implemented.

The simulations used to calculate control values were implemented using the bioLQM toolkit developed by the CoLoMoTo Consortium \cite{Naldi2}. Using bioLQM, we implement our Boolean models of the networks, find their attractors, and simulate the system's trajectories using the random asynchronous update mode.

The logistic regressions were created and analyzed using the scikit-learn module. From the linear\_model submodule, the LogisticRegression function is used to generate a fit of the data. This function is implemented using the liblinear solver and a regularization strength of 100. The scikit-learn module was also used to obtain the precision-recall curve (\texttt{metrics.precision\_recall\_curve}) and to calculate the AUPRC (\texttt{metrics.average\_precision\_score}).

Bootstrapping was used to approximate the precision of the set of all node subsets from random samples of these subsets. Bootstrapping is a resampling method that generates a distribution from a sample set of subsets to make inferences about the value of an observable (precision in our case) in the set of all node subsets. The sample sets were randomly chosen samples of subsets. We tested two separate sets: the set of every node subset and the set of FVS subsets. For 1 node subsets, the random samples chosen are of size 25 or include all 1 node subsets if there are less than 25 total node subsets. For 2 and 3 node subsets, 100 subsets are chosen for the random sample of all node subsets, and 50 subsets are chosen for the random sample of FVS subsets. After picking a random sample, every subset in the sample was simulated to determine its \textit{To Control} and \textit{Away Control} value. For each random sample, we resample with replacement a resampled set of the same size. We resample 1000 times and calculate the precision of each resampled set. From this distribution of these 1000 precision values, we can approximate the precision value of the entire set of node subsets or FVS subsets depending on the original set that the random sample is generated from.

\begin{acknowledgments}
\textbf{Funding:} This work was supported by NSF grants IIS 1814405, MCB 1715826.
\textbf{Author Contributions:} J.G.T.Z. conceptualized the idea and J.G.T.Z. and E.N. developed the methodology with input from R.A.. R.A. acquired funding and provided project administration. E.N. programmed the software with help from J.G.T.Z.. E.N. investigated the problem through conducting simulations, and provided formal analysis of the resulting data under the supervision of both R.A. and J.G.T.Z.. E.N. visualized the data and wrote the original draft, and J.G.T.Z. and R.A. reviewed and edited the final work. J.G.T.Z. and R.A. contributed equally to this work as senior authors.
\end{acknowledgments}

\bibliography{References}

\setcounter{table}{0}
\renewcommand{\thetable}{S\arabic{table}}
\renewcommand{\thefigure}{S\arabic{figure}}
\newpage
\widetext
\begin{center}
\textbf{\large Supplemental Materials}
\end{center}

\begin{table*}[ht]
\caption{\label{tabS1} Average rank of the AUPRC values for each network for FVS subsets of size one to three among the seven topological metrics and three intersection metrics. For each metric, we present the average rank, from 1 to 10, of the AUPRC values for both the \textit{To Control} and \textit{Away Control} values. The intersection metrics and three propagation topological metrics perform similarly, and better than the other four metrics.}
\begin{ruledtabular}
\begin{tabular}{l*{2}{c}}
&\textit{To Control}&\textit{Away Control}\\
\hline
All Metrics&4.38&4.75\\
Propagation Metrics&5.83&5.21\\
Modified PRINCE and CheiRank&4.67&3.71\\
PRINCE&5.71&5.71\\
Modified PRINCE&4.04&3.42\\
CheiRank&4.83&3.88\\
Out-degree&6.75&5\\
Average Distance&6.42&6.75\\
Positive Cycles&6.25&7.46\\
SCC Size&6.13&9.13
\end{tabular}
\end{ruledtabular}
\end{table*}

\begin{table*}
\caption{\label{tabS2}Percentile cutoff needed for a precision of 0.95 for four networks and FVS subsets of sizes 1, 2, and 3. Starting at the percentile cutoff of 0, which is the least restrictive cutoff, we find the precision of the identified subsets and increase the cutoff until this precision goes above 0.95. As more metrics are added to the intersection, this value decreases, indicating that a less restrictive cutoff is needed to find subsets that have high control values. }
\begin{ruledtabular}
\begin{tabular}{ll*{6}{c}}
&&\multicolumn{2}{c}{1 Nodes}&\multicolumn{2}{c}{2 Nodes}&\multicolumn{2}{c}{3 Nodes}\\
\cline{3-4}\cline{5-6}\cline{7-8}
&&\textit{To}&\textit{Away}&\textit{To}&\textit{Away}&\textit{To}&\textit{Away}\\
\hline
T-LGL&All Metrics&48&65&87&42&91&24\\
&Propagation Metrics&53&65&87&57&94&28\\
&Modified PRINCE + CheiRank&53&89&91&77&96&53\\
&CheiRank&53&89&94&80&98&53\\
NSCLC&All Metrics&60&60&80&80&89&84\\
&Propagation Metrics&80&90&88&83&86&76\\
&Modified PRINCE + CheiRank&80&90&88&84&87&78\\
&CheiRank&95&95&97&97&92&88\\
FA/BRCA Variant \#1&All Metrics&100&44&100&38&100&16\\
&Propagation Metrics&100&56&100&38&100&11\\
&Modified PRINCE + CheiRank&100&56&100&32&100&11\\
&CheiRank&100&56&100&32&99&11\\
Helper T Cells&All Metrics&25&33&26&64&18&23\\
&Propagation Metrics&25&75&54&74&66&66\\
&Modified PRINCE + CheiRank&25&83&54&78&78&81\\
&CheiRank&25&100&54&66&81&81
\end{tabular}
\end{ruledtabular}
\end{table*}

\begin{table*}
\caption{\label{tabS3}.Number of FVS subsets above the percentile cutoff needed for a precision of 0.95 for the four networks and FVS subsets of sizes 1, 2, and 3. Overall, the propagation intersection metric identifies the most subsets. Among all of the networks and FVS subsets sizes, the average rank of the propagation intersection among the three intersection metrics and seven topological metrics was 2.83, with 1 being the most FVS subsets identified and 10 being the least FVS subsets identified. This is the lowest average rank, with the Modified PRINCE being the second lowest (2.96), and the Modified PRINCE and CheiRank intersection metric being the third lowest (3.33). The all metric intersection metric had an average rank of 3.96, which is the 5th lowest average rank, so the all metric intersection metric performs better than the non-propagation topological metrics, but is not as effective as an intersection of a subset of the metrics it contains. This shows that using intersections may increase the number of subsets, but as more metrics are added to the intersection, it is possible for the intersection metric to be too strict and will cause the intersection metric to be less predictive at identifying successful FVS subsets.}
\begin{ruledtabular}
\begin{tabular}{ll*{6}{c}}
&&\multicolumn{2}{c}{1 Nodes}&\multicolumn{2}{c}{2 Nodes}&\multicolumn{2}{c}{3 Nodes}\\
\cline{3-4}\cline{5-6}\cline{7-8}
&&\textit{To}&\textit{Away}&\textit{To}&\textit{Away}&\textit{To}&\textit{Away}\\
\hline
T-LGL&All Metrics&1&2&1&42&8&301\\
&Propagation Metrics&1&5&4&42&13&362\\
&Modified PRINCE + CheiRank&1&2&2&25&13&259\\
&CheiRank&1&2&2&27&7&280\\
NSCLC&All Metrics&2&2&6&6&7&10\\
&Propagation Metrics&2&1&11&16&67&144\\
&Modified PRINCE + CheiRank&2&1&15&21&91&180\\
&CheiRank&1&1&6&6&91&131\\
FA/BRCA Variant \#1&All Metrics&0&2&0&11&0&48\\
&Propagation Metrics&0&3&0&14&0&62\\
&Modified PRINCE + CheiRank&0&3&0&21&0&64\\
&CheiRank&0&4&0&24&1&68\\
Helper T Cells&All Metrics&5&4&21&1&86&60\\
&Propagation Metrics&5&2&15&12&40&43\\
&Modified PRINCE + CheiRank&5&2&19&12&32&29\\
&CheiRank&5&0&24&22&36&38
\end{tabular}
\end{ruledtabular}
\end{table*}

\begin{table*}
\renewcommand{\arraystretch}{0.5}
\caption{\label{tabS4}Description of key properties of the networks analyzed in this work. For each network, we indicate the number of nodes, edges, attractors, and the size of the minimal FVS. We also indicate the source node values and the mutations used for each network. The number of attractors in the system are based on the specified source node values and mutations.}
\begin{ruledtabular}
\begin{tabular}{*{7}{c}}
Network&\# Nodes&\# Edges&\# Attractors&\shortstack{Size of\\Minimal\\FVS}&\shortstack{Source\\Node\\Values}&Mutation\\
\hline
\\
T-LGL&60&141&3&12&CD45=0&N/A\\
&&&&&IL15=1&\\
&&&&&PDGF=0&\\
&&&&&Stimuli=1&\\
&&&&&Stimuli2=0&\\
&&&&&TAX=0&\\
\\
NSCLC&33&361&5&17&N/A&N/A\\
\\
Helper T Cell&36&108&12&10&APC=1&N/A\\
&&&&&IFNB\_e=0&\\
&&&&&IFNG\_e=0&\\
&&&&&IL10\_e=0&\\
&&&&&IL12\_e=0&\\
&&&&&IL15\_e=0&\\
&&&&&IL21\_e=0&\\
&&&&&IL23\_e=0&\\
&&&&&IL27\_e=0&\\
&&&&&IL2\_e=1&\\
&&&&&IL4\_e=0&\\
&&&&&IL6\_e=0&\\
&&&&&TGFB\_e=1&\\
\shortstack{FA/BRCA\\Variant \#1}&28&121&2&8&N/A&MUS81=0\\
\shortstack{FA/BRCA\\Variant \#2}&28&116&2&8&N/A&MRN=0\\
\shortstack{FA/BRCA\\Variant \#3}&28&117&2&8&N/A&BRCA1=0\\
\\
Geroconversion&26&67&2&4&Insulin=1&N/A\\
&&&&&GF=1&\\
&&&&&Therapy=0&\\
\shortstack{MAPK\\Variant \#1}&53&100&2&6&N/A&EGFR=1\\
\shortstack{MAPK\\Variant \#2}&53&101&3&6&N/A&FGFR3=1
\end{tabular}
\end{ruledtabular}
\end{table*}
\clearpage

\begin{table*}
\caption{\label{tabS5}Excel file containing the precision and recall of the identified subsets for the first array of four networks: the T-LGL, NSCLC, first variant of FA/BRCA, and Helper T Cell networks. For each network, we identified FVS subsets based on the all metric intersection metric and the propagation intersection metric. For each intersection metric, we identify the FVS subsets in one of four ways. 1) FVS subsets with a percentile cutoff greater than the maximum value of the 0.9 precision crossing points; 93/90 for \textit{To Control} and \textit{Away Control} respectively. 2) FVS subsets with a percentile cutoff greater than the third quartile value of the 0.9 precision crossing points; 83/72.75 for \textit{To Control} and \textit{Away Control} respectively. 3) FVS subsets predicted as successful by an unbalanced Logistic Regression. 4) FVS subsets predicted as successful by a balanced Logistic Regression. For each identification method, we recorded the precision and recall of the identified subsets for subsets of size 1, 2, and 3 for both \textit{To Control} and \textit{Away Control} on all four networks.}
\end{table*}

\begin{table*}
\caption{\label{tabS6}Excel file containing the precision of the identified subsets for the seconds array of five networks: the Geroconversion, second and third variants of the FA/BRCA, and two variants of the MAPK networks. For these networks, we identified FVS subsets using only the two percentile cutoff identification methods from Table \ref{tabS5}. Using the two percentile cutoff identification methods, we identify FVS subsets of size 1, 2, and 3 for both \textit{To Control} and \textit{Away Control} on all five networks. We then measure the precision of the identified set of FVS subsets by finding the percent of identified subsets that can achieve a control value above 0.9.}
\end{table*}
\end{document}